\documentclass[12pt]{article}

\usepackage{amsmath,amssymb,amsfonts,amsthm,hyperref}
\usepackage{graphicx,multirow,caption,bbm,latexsym,epsfig}
\usepackage{cite}

\newcommand{\bs}{\begin{subequations}}
\newcommand{\es}{\end{subequations}}
\newcommand{\be}{\begin{equation}}
\newcommand{\ee}{\end{equation}}
\newcommand{\ba}{\begin{eqnarray}}
\newcommand{\ea}{\end{eqnarray}}
\newcommand{\no}{\nonumber \\}
\newcommand{\mc}{\mathcal{M}}
\newcommand{\diag}{\mbox{diag}}
\newcommand{\bone}{\mathbbm{1}}

\allowdisplaybreaks

\begin{document}

\title{
\normalsize \hfill CFTP/20-008 \\
\hfill UWThPh-2020-23
\\[3mm]
\LARGE Radiative seesaw corrections and charged-lepton decays \\
in a model with soft flavour violation}

\author{
  E.~H.~Aeikens,$^{(1)}$\thanks{E-mail: \tt elke.aeikens@univie.ac.at}\
  \addtocounter{footnote}{1}
  P.~M.~Ferreira,$^{(2,3)}$\thanks{E-mail: \tt pmmferreira@fc.ul.pt}\ \
  W.~Grimus,$^{(1)}$\thanks{E-mail: \tt walter.grimus@univie.ac.at}\ \
  D.~Jur\v{c}iukonis,$^{(4)}$\thanks{E-mail:
    \tt darius.jurciukonis@tfai.vu.lt}
  \\*[1mm]
  and L.~Lavoura$^{(5)}$\thanks{E-mail: \tt balio@cftp.tecnico.ulisboa.pt}
  \\*[2mm]
  $^{(1)}\!$
  \small University of Vienna, Faculty of Physics,
  \small Boltzmanngasse~5, A-1090 Wien, Austria
  \\*[2mm]
  $^{(2)}\!$
  \small Instituto~Superior~de~Engenharia~de~Lisboa~---~ISEL,
  \small 1959-007~Lisboa, Portugal
  \\*[2mm]
  $^{(3)}\!$
  \small Centro~de~F\'\i sica~Te\'orica~e~Computacional,
  \small Faculdade~de~Ci\^encias, Universidade~de~Lisboa, \\
  \small Av.~Prof.~Gama~Pinto~2, 1649-003~Lisboa, Portugal
  \\*[2mm]
  $^{(4)}\!$
  \small Vilnius University, Institute of Theoretical Physics and Astronomy, \\
  \small Saul\.etekio~ave.~3, Vilnius 10257, Lithuania
  \\*[2mm]
  $^{(5)}\!$
  \small Universidade de Lisboa, Instituto Superior T\'ecnico, CFTP, \\
  \small Av.~Rovisco~Pais~1, 1049-001~Lisboa, Portugal
  \\*[2mm]
}

%%%%% LAST CHANGES MADE BY LUIS
%%%%% \date{September 28, 2020}
\date{November 2, 2020}

\maketitle

\begin{abstract}
  We consider the one-loop radiative corrections
  to the light-neutrino mass matrix and their consequences
  for the predicted branching ratios
  of the five lepton-flavour-violating decays
  $\ell_1^- \to \ell_2^- \ell_3^+ \ell_3^-$
  in a two-Higgs-doublet model
  furnished with the type-I seesaw mechanism
%%%%%  and \emph{soft} left-flavour violation.
  and \emph{soft} lepton-flavour violation.
  We find that the radiative corrections are very significant;
  they may alter the predicted branching ratios
  by several orders of magnitude and,
  in particular,
  they may help explain why
  $\mbox{BR} \left( \mu^- \to e^- e^+ e^- \right)$
  is  strongly suppressed relative to the branching ratios
  of the decays of the $\tau^-$.
  We conclude that, 
  in any serious numerical assessment of the predictions of this model,
  it is absolutely necessary to take into account
  the one-loop radiative corrections to the light-neutrino mass matrix.

\end{abstract}

\newpage

\section{Introduction}
\label{introduction}

The existence of neutrino oscillations
is now firmly established---see~\cite{nobel1,nobel2,nobel3,rpp}
and references therein.
Therefore,
the violation of the family lepton numbers $L_\ell$
($\ell = e,\mu,\tau$) 
is firmly established as well.
However,
no violation of the $L_\ell$ but for neutrino oscillations
has been hitherto detected.
In this context,
the flavour-violating charged-lepton decays are of particular importance,
because it is expected that in the near future
the experimental bounds on the branching ratios (BRs)
of those decays will be improved
substantially~\cite{Mu3e,aushev,cerri,abdul,Belle-II}
(see also section~2 of~\cite{vicente}). 
It is thus important to address those decays in specific models
for the neutrino masses and lepton mixings---and the more so since,
when one incorporates neutrino masses and lepton mixings
in the Standard Model (SM),
those BRs are so small that the decays
%%%%% are in practice invisible~\cite{petcov,bilenky,blackstone}.
are in practice invisible~\cite{petcov,bilenky,roig,blackstone}.

%%%%% I HAVE REWRITTEN THE FOLLOWING TWO PARAGRAPHS.
In this letter we discuss the model put forward in~\cite{GL1,aeikens}.
This is in general \emph{a multi-Higgs-doublet extension of the SM}
(for reviews see~\cite{review,ivanov}),
but we confine ourselves to just two Higgs doublets.
The model has \emph{three right-handed neutrino singlets $\nu_{\ell R}$
  that enable the seesaw mechanism}~\cite{seesaw1,seesaw2,seesaw3,
  seesaw4,seesaw5}.
The lepton Yukawa couplings are
\ba
\label{yukawas1}
\mathcal{L}_\mathrm{Yukawa} &=& - \sum_{\ell_1,\, \ell_2 = e, \mu, \tau} \left[
  \left( \begin{array}{cc} \varphi_1^-, & {\varphi_1^0}^\ast \end{array} \right)
  \bar \ell_{1R} \left( Y_1 \right)_{\ell_1 \ell_2}
  + \left( \begin{array}{cc} \varphi_1^0, & - \varphi_1^+ \end{array} \right)
  \bar\nu_{\ell_1 R} \left( Z_1 \right)_{\ell_1 \ell_2}
  \right. \no & & \left.
  + \left( \begin{array}{cc} \varphi_2^-, & {\varphi_2^0}^\ast
  \end{array} \right) \bar\ell_{1R} \left( Y_2 \right)_{\ell_1 \ell_2}
  + \left( \begin{array}{cc} \varphi_2^0, & - \varphi_2^+ \end{array} \right)
  \bar\nu_{\ell_1 R} \left( Z_2 \right)_{\ell_1 \ell_2}
  \right] \left( \begin{array}{c} \nu_{\ell_2 L} \\
  \ell_{2L} \end{array} \right)
  + \mathrm{H.c.}, \hspace*{2mm}
\ea
where $Y_{1,2}$ and $Z_{1,2}$ are Yukawa-coupling matrices.
A crucial feature of the model is
\emph{the imposition of three global $U(1)_\ell$
symmetries associated with the family lepton numbers $L_\ell$};
those symmetries force $Y_{1,2}$ and $Z_{1,2}$ to be diagonal.
(Naturally,
the lepton numbers of the two Higgs doublets are zero.)
Without loss of generality,
in this letter we use the `Higgs basis',
wherein only the first doublet
has a nonzero vacuum expectation value $v \left/ \sqrt{2} \right.$,
where $v \simeq 246$\,GeV is real and positive,
in its neutral component $\varphi_1^0$.
This allows us to rewrite~\eqref{yukawas1} as
\ba
\label{yukawas}
\mathcal{L}_\mathrm{Yukawa} &=& - \sum_{\ell = e, \mu, \tau} \left[
  \left( \begin{array}{cc} \varphi_1^-, & {\varphi_1^0}^\ast \end{array} \right)
  \bar\ell_R\, \frac{\sqrt{2} m_\ell}{v} 
  + \left( \begin{array}{cc} \varphi_1^0, & - \varphi_1^+ \end{array} \right)
  \bar\nu_{\ell R}\, d_\ell
  \right. \no & & \left.
  + \left( \begin{array}{cc} \varphi_2^-, & {\varphi_2^0}^\ast
  \end{array} \right) \bar\ell_R\, \gamma_\ell
  + \left( \begin{array}{cc} \varphi_2^0, & - \varphi_2^+ \end{array} \right)
  \bar\nu_{\ell R}\, \delta_\ell
  \vphantom{\frac{\sqrt{2} m_\ell}{v}}
  \right] \left( \begin{array}{c} \nu_{\ell L} \\ \ell_L \end{array} \right)
  + \mathrm{H.c.},
\ea
where the $m_\ell$ are the (real and positive) charged-lepton masses and
$d_\ell$,
$\gamma_\ell$,
and $\delta_\ell$ are dimensionless and,
in general,
complex Yukawa coupling constants.

In our model \emph{the source of lepton-flavour violation lies exclusively
in the Majorana mass matrix $M_R$ of the right-handed neutrinos}.
In other words,
the only lepton-flavour-violating (LFV) terms in the Lagrangian are in
\be\label{Lmass}
\mathcal{L}_{\nu_R\, \mathrm{mass}} = - \frac{1}{2}
\sum_{\ell_1, \ell_2 = e, \mu, \tau}\!
\left( M_R \right)_{\ell_1 \ell_2}
\bar\nu_{\ell_1 R}\, C\, \bar\nu_{\ell_2 R}^T
+ \mathrm{H.c.},
\ee
where $\left( M_R \right)_{\ell_1 \ell_2} = \left( M_R \right)_{\ell_2 \ell_1}$
are coefficients with mass dimension
and $C$ is the charge-conjugation matrix in Dirac space.
The salient feature of this model is the \emph{soft} nature of the breaking
of the $L_\ell$~\cite{GL1,soft} 
by $\mathcal{L}_{\nu_R\, \mathrm{mass}}$.
Soft-breaking of a symmetry means that the symmetry is preserved
by dimension-four terms in the Lagrangian,
\textit{viz.}\ the Yukawa couplings~\eqref{yukawas},
but it is broken by terms with mass dimension smaller than four,
\textit{viz.}\ the Majorana masses~\eqref{Lmass}.
In our model,
the softness of the breaking
ensures that the one-loop amplitudes of LFV charged-lepton decays are finite,
as was explicitly demonstrated by two of us
in~\cite{GL1}.\footnote{This mechanism for suppressing
  undesirable flavour-changing neutral currents
  has no counterpart in the quark sector.
  Since we do not want to set to zero
  any of the Yukawa couplings in~\eqref{yukawas},
  the two Higgs doublets cannot transform non-trivially
  under any global symmetry. 
  Therefore,
  at this stage,
  we can only resort to finetuning in the quark sector.
  In the present letter we shall not address
  this issue any further.}

Let $S^\pm_a$ ($a = 1, 2$) and $S^0_b$ ($b = 1, 2, 3, 4$) denote,
respectively,
the charged-scalar and the (real) neutral-scalar
mass eigenfields of our two-Higgs-doublet model (2HDM).
By definition,
$S^+_1 \equiv G^+$ and $S^0_1 \equiv G^0$ are,
respectively,
the charged and the neutral Goldstone bosons.
Again by definition,
$S^0_2 \equiv H$ is the physical scalar with mass $m_H \simeq 125$\,GeV 
that has been observed at the LHC.
Let $M_3$ and $M_4$ denote the masses of $S^0_3$ and $S^0_4$, 
respectively.

It is an outstanding feature of our model that
the amplitudes for the radiative decays $\ell_1^\pm \to \ell_2^\pm \gamma$
and $Z^0 \to \ell_1^+ \ell_2^-$ ($\ell_1 \neq \ell_2$)
are suppressed by $m_R^{-2}$,
where $m_R$ is the seesaw scale~\cite{GL1};
one can estimate that for $m_R \gtrsim 10^3$\,TeV these decays are invisible,
in the foreseeable future,
in the context of our model~\cite{aeikens}
(another model with this feature is discussed in~\cite{chowdhury}).
The same suppression occurs when the gauge bosons
are off-mass shell,
\textit{viz.}\ in the one-loop diagrams
for the LFV decays $\mu^- \to e^- e^+ e^-$
and $\tau^- \to \ell_2^- \ell_3^+ \ell_3^-$
($\ell_2, \ell_3 = e, \mu$)~\cite{GL1}
where those decays are mediated by either a virtual $\gamma$
or a virtual $Z^0$.\footnote{The box diagrams
  for $\mu^- \to e^- e^+ e^-$ and $\tau^- \to \ell_2^- \ell_3^+ \ell_3^-$
  are also suppressed by $1/m_R^2$~\cite{GL1}.}
On the other hand,
those five decays also have one-loop amplitudes
mediated by neutral-scalar exchange, and
these amplitudes are \emph{unsuppressed} when $m_R \to \infty$~\cite{GL1}. 
It is the purpose of this letter to present a theoretical and numerical study
of these three-body decays while
\emph{taking into account the radiative corrections 
to the seesaw mass matrix of the light neutrinos}~\cite{GL2}. 
The latter point is new when compared to~\cite{aeikens},
and it is important because of two reasons:
\begin{itemize}
\item For $M_3$ or $M_4$ larger than $4 \pi v \sim 3$\,TeV,
  and provided the Yukawa couplings $d_\ell$ and $\delta_{\ell'}$
  are of similar order of magnitude,
  the radiative corrections to the neutrino mass matrix
  are dominant;\footnote{It was already
    stressed in~\cite{neufeld}
    (see also~\cite{ibarra,jurciukonis})
    that the radiative corrections to the seesaw mechanism,
    in the presence of two or more Higgs doublets and heavy neutral scalars,
    may be quite large.
    Moreover,
    it has been demonstrated in~\cite{aristizabal} that,
    even with only one Higgs doublet,
    those corrections may be substantial
    for fine-tuned tree-level neutrino mass matrices.}
  they are non-negligible even for values of $M_{3,4}$
  much lower than that.
\item The branching ratios
  $\mbox{BR} \left( \ell_1^- \to \ell_2^- \ell_3^+ \ell_3^- \right)$
  depend on the mass matrix
  $M_R$~\cite{GL1,aeikens}---see section~\ref{light}.
  Information on $M_R$ is not directly available
  but has to be extracted from the mass matrix of the light neutrinos.
  The latter matrix may be assembled from the light-neutrino masses
  and from the lepton mixing obtained from fits
  to the neutrino oscillation data.
  Therefore,
  the radiative corrections to the mass matrix of the light neutrinos
  will influence the extraction of $M_R$.
  Indeed,
  they may alter the branching ratios drastically,
  as we shall see later.
\end{itemize}
Henceforth,
for the sake of brevity,
the acronym `BR' will always refer to the branching ratios
of the five decays $\ell_1^- \to \ell_2^- \ell_3^+ \ell_3^-$;
the same will apply to the phrase `decay rate'.

Although in this letter we consider just a 2HDM,
we nevertheless have a large number of parameters. 
In order to facilitate the numerical analysis
it is useful to reduce that number.
We adopt the strategy
of~\cite{aeikens} and assume the following: 
\begin{enumerate}
\renewcommand{\labelenumi}{\Alph{enumi}.}
\item There is no mixing between the two scalar doublets.
\item All parameters are real.
\end{enumerate}
Through assumption~A,
the mixing of the scalars
is simplified to~\cite{aeikens}
\be
\label{doublets}
\varphi_1^+ = G^+, \quad \varphi_2^+ = H^+, \quad
\varphi_1^0 = \frac{v + H + i G^0}{\sqrt{2}}, \quad
\varphi_2^0 = e^{-i\alpha}\, \frac{S_3^0 + i S_4^0}{\sqrt{2}},
\ee
where $H^+$ is a physical charged scalar which,
however,
plays no role in this letter.
The advantage of assumption~A is threefold:
\begin{itemize}
\item There are in general three parameters
  in the mixing of the neutral scalars~\cite{aeikens,osland}.
  With assumption~A they
  are reduced to only one---the phase $\alpha$,
  which is,
  however,
  unphysical because one may freely rephase $\varphi_2^+$ and
  $\varphi_2^0$.\footnote{Due to assumption~B,
    later on we will set $e^{-i\alpha} = 1$ in~\eqref{doublets}.}
\item The formulas for the BRs simplify considerably
  (see section~\ref{branching}).
\item The couplings of $H$ are identical to the ones in the SM,
  hence the experimental restrictions on the couplings of $H$
  are automatically fulfilled.
\end{itemize}
We thus consider that we are in the exact \textit{alignment limit}\/
of the 2HDM.
This is in accordance with the measurements of the properties
of the scalar discovered at LHC (see for instance~\cite{aad1,aad2}),
which have found that that scalar behaves in a manner very similar
to the SM Higgs boson;
its couplings are experimentally constrained to be very close
to their respective SM values.
Assumption~A ensures that this indeed happens in our 2HDM.
In the ensuing discussions
we will initially keep complex parameters,
but we take into account assumption~A right from the beginning.

This letter is organized as follows.
In section~\ref{light} 
we discuss the light-neutrino mass matrix,
including the radiative corrections.
The formulas for the decay rates are displayed in section~\ref{branching},
where we also derive a prediction of our model when assumption~A holds.
In section~\ref{numerical} we discuss the procedure
of our numerical investigation and in
section~\ref{results} some results thereof are presented.
We draw our conclusions in section~\ref{conclusions}.
An appendix makes a digression through the scalar potential of the 2HDM
in order to demonstrate that the two new scalars $S^0_3$ and $S^0_4$
may have sufficiently different masses.

\section{The light-neutrino mass matrix}
\label{light}

The Majorana mass matrix $\mc_\nu$ of the light neutrinos is diagonalized as
\be
\label{Mdiag}
U_L^T \mc_\nu U_L = \hat m \equiv \diag\left( m_1, m_2, m_3 \right),
\ee
where $U_L$ is $3 \times 3$ unitary
and the $m_j$ ($j = 1, 2, 3$) are real and non-negative.
Since the charged-lepton mass matrix
is diagonal from the start, \textit{cf.}~\eqref{yukawas},
$U_L$ is just the lepton mixing matrix.
The matrix $M_R$,
defined in~\eqref{Lmass},
is diagonalized as
\be
\label{UR}
U_R^\dagger M_R U_R^\ast = \widetilde m
\equiv \diag \left( m_4, m_5, m_6 \right),
\ee
where the $m_{3+j}$ are real and positive
and the matrix $U_R$ is $3 \times 3$ unitary.
For the decay rates---see~\eqref{fjfio} in the next section---we need
the quantities~\cite{GL1,aeikens}
\be\label{X}
X_{\ell_1 \ell_2} = \frac{1}{16 \sqrt{2} \pi^2}\,
\sum_{j=1}^3 \left( U_R \right)_{\ell_1 j} \left( U_R^\ast \right)_{\ell_2 j}\,
%%%%% \ln{m_{3+j}^2}\,
\ln{\frac{m_{3+j}^2}{\mu^2}}
\ee
%
%%%%% for $\ell_1 \neq \ell_2$.\footnote{There is no need to write
%%%%%   in~\eqref{X} an
%%%%%   (arbitrary)
%%%%%   renormalization scale to render the argument
%%%%%   of the logarithm dimensionless,
for $\ell_1 \neq \ell_2$.\footnote{The renormalization scale $\mu$
  that renders the argument of the logarithm dimensionless is arbitrary,
  since $U_R$ is unitary and only the case $l_1 \neq l_2$
  is considered in this work.}
This requires us to know both $U_R$ and
the heavy-neutrino masses
$m_{4,5,6}$.
Note that,
since $U_R$ is a $3 \times 3$ unitary matrix,
$X_{\ell_1 \ell_2}$ cannot be large;
one has $\left| X_{\ell_1 \ell_2} \right| \lesssim 0.1$
if $10^9\, \mathrm{GeV} \le m_{4,5,6} \le 10^{19}$\,GeV.

We parameterize $U_L$ as
\bs
\label{UL}
\ba
\label{PMNSdef}
U_L &=& e^{i \hat \alpha}\, U_\mathrm{PMNS}\, e^{i \hat \beta},
\\*[2mm]
\label{PMNS}
U_\mathrm{PMNS} &=& \left( \begin{array}{ccc}
  c_{12} c_{13} & s_{12} c_{13} & \epsilon^\ast \\
  - s_{12} c_{23} - c_{12} s_{23}\, \epsilon &
  c_{12} c_{23} - s_{12} s_{23}\, \epsilon & s_{23} c_{13} \\
  s_{12} s_{23} - c_{12} c_{23}\, \epsilon &
  - c_{12} s_{23} - s_{12} c_{23}\, \epsilon & c_{23} c_{13}
\end{array} \right),
\\
\label{jvugi}
\epsilon &\equiv& s_{13} \exp{\left( i \delta \right)}.
\ea
\es
In~\eqref{PMNSdef},
$e^{i \hat \alpha}$ and $e^{i \hat \beta}$ are diagonal matrices of phase factors
while $U_\mathrm{PMNS}$ is the
Pontecorvo--Maki--Nakagawa--Sakata matrix~\cite{rpp}.
Out of the three phases in $e^{i \hat \beta}$,
one may be absorbed into $\hat \alpha$
and the remaining two are the so-called Majorana phases,
which are physically meaningful quantities.
In~\eqref{PMNS} and~\eqref{jvugi},
$c_{ij} = \cos{\theta_{ij}}$ and $s_{ij} = \sin{\theta_{ij}}$
%%%%% for $ij = 12, 13, 23$.
for $ij = 12, 13, 23$,
and $\delta$ is a $CP$-violating phase.
The matrix $U_R$ may be parameterized in the same way as $U_L$.

The matrix $\mc_\nu$ is the sum of two parts:
\be
\mc_\nu = M_\nu^\mathrm{tree} + \delta M_L,
\ee
where the tree-level part $M_\nu^\mathrm{tree}$ is given by the seesaw mechanism
and the one-loop-level part $\delta M_L$
is generated by the
radiative corrections.
As is well known,
\be
\label{vfpro}
M_\nu^\mathrm{tree} = - M_D^T M_R^{-1} M_D,
\ee
where $M_D$ is the neutrino Dirac mass matrix.
Referring to~\eqref{yukawas},
let us define the diagonal matrices
\be
%%%%% NEW LABEL
\label{Delta1Delta2}
\Delta_1 = \diag \left( d_e,\, d_\mu,\, d_\tau \right)
\quad \mbox{and} \quad
\Delta_2 = \diag \left( \delta_e,\, \delta_\mu,\, \delta_\tau \right).
\ee
Because we use the Higgs basis,
$M_D$ is given by 
\be
M_D = \frac{v}{\sqrt{2}} \, \Delta_1,
\ee
hence it is diagonal.
Thus,
\be
\label{tree2}
M_\nu^\mathrm{tree} = - \frac{v^2}{2}\, \Delta_1\, U_R^\ast\,
\frac{1}{\widetilde m}\,
U_R^\dagger\, \Delta_1.
\ee
The radiative part of $\mc_\nu$ is given by~\cite{GL2}
\bs\label{uit}
\ba
\delta M_L &=&
\frac{3 m_Z^2}{32 \pi^2}\, \Delta_1\, U_R^\ast \left(
\frac{1}{\widetilde{m}}
  \ln{\frac{{\widetilde m}^2}{m_Z^2}} \right) U_R^\dagger\, \Delta_1
\label{uit1} \\ & &
+ \frac{m_H^2}{32 \pi^2}\, \Delta_1\, U_R^\ast \left( \frac{1}{\widetilde{m}}
\ln{\frac{{\widetilde m}^2}{m_H^2}} \right) U_R^\dagger\, \Delta_1
\label{uit2} \\ & &
+ \frac{M_3^2}{32 \pi^2}\,
e^{- 2 i \alpha} \Delta_2\, U_R^\ast \left( \frac{1}{\widetilde{m}}
\ln{\frac{{\widetilde m}^2}{M_3^2}} \right) U_R^\dagger\, \Delta_2
\label{uit3} \\ &&
- \frac{M_4^2}{32 \pi^2}\,
e^{- 2 i \alpha} \Delta_2\, U_R^\ast \left( \frac{1}{\widetilde{m}}
\ln{\frac{{\widetilde m}^2}{M_4^2}} \right) U_R^\dagger\, \Delta_2,
\label{uit4}
\ea
\es
where the four lines correspond successively to the contribution
of the $Z^0$ gauge boson with mass $m_Z$,
of the SM scalar $H$,
and of the new scalars $S^0_3$ and $S^0_4$.
In~\eqref{uit}
we have already taken into account assumption~A of section~\ref{introduction}.
We have also used $m_{4,5,6} \gg m_Z, m_H, M_{3,4}$.
It is clear that for $M_{3,4} \gtrsim 4 \pi v$
and provided $\Delta_1$ and $\Delta_2$ are of identical orders of magnitude,
the contributions~\eqref{uit3} and~\eqref{uit4} dominate
%%%%% over the contribution~\eqref{tree2}.\footnote{The
over the contribution~\eqref{tree2} provided $M_3 \neq M_4$.\footnote{The
  large logarithms of $m_{3+j} \left/ M_{3,4} \right.$ further enhance
  the contributions~\eqref{uit3} and~\eqref{uit4}.}
Lines~\eqref{uit3} and~\eqref{uit4}
coincide with the well-known scotogenic mechanism;
however, 
in the scotogenic model proper~\cite{ma}
the Yukawa couplings $d_\ell$ and $\gamma_\ell$
are zero (because of an additional symmetry),
while in this letter they are nonzero.

We reformulate~\eqref{Mdiag} to
\bs
\label{uit5}
\ba
\label{tree}
e^{- i \hat \alpha}\, U_\mathrm{PMNS}^\ast
\left( \hat m\, e^{- 2 i \hat \beta} \right)
U_\mathrm{PMNS}^\dagger\, e^{- i \hat \alpha} &=&
- \frac{v^2}{2}\, \Delta_1\, U_R^\ast\,
\frac{1}{\widetilde{m}}\, U_R^\dagger\,
\Delta_1
\\ \label{Z} & &
+ \frac{3 m_Z^2}{32 \pi^2}\, \Delta_1\, U_R^\ast \left(
\frac{1}{\widetilde{m}}
  \ln{\frac{{\widetilde m}^2}{m_Z^2}} \right) U_R^\dagger\, \Delta_1
\\ \label{H} & &
+ \frac{m_H^2}{32 \pi^2}\, \Delta_1\, U_R^\ast \left( \frac{1}{\widetilde{m}}
\ln{\frac{{\widetilde m}^2}{m_H^2}} \right) U_R^\dagger\, \Delta_1
\\ \label{3} & &
+ \frac{M_3^2}{32 \pi^2}\,
e^{- 2 i \alpha} \Delta_2\, U_R^\ast \left( \frac{1}{\widetilde{m}}
\ln{\frac{{\widetilde m}^2}{M_3^2}} \right) U_R^\dagger\, \Delta_2
\\ \label{4} & &
- \frac{M_4^2}{32 \pi^2}\,
e^{- 2 i \alpha} \Delta_2\, U_R^\ast \left( \frac{1}{\widetilde{m}}
\ln{\frac{{\widetilde m}^2}{M_4^2}} \right) U_R^\dagger\, \Delta_2.
\ea
\es
Equation~\eqref{uit5} is the basis for our numerical computations.

The diagonal matrix $e^{i\hat \alpha}$
in the left-hand side of~\eqref{uit5} is irrelevant;
indeed,
it can be absorbed into $U_R$
in the right-hand side,
since $\Delta_1$ and $\Delta_2$ are diagonal matrices.
In principle we use as input the Majorana phases,
$U_\mathrm{PMNS}$,
$\hat{m}$,
$\Delta_1$,
$\Delta_2$,
$\alpha$,
$M_3$,
and $M_4$
(and additionally the fixed values $v = 246$\,GeV, 
$m_Z = 91$\,GeV,
and $m_H = 125$\,GeV)
and we solve~\eqref{uit5} to find
the three $m_{3+j}$ and the nine parameters
of the $3 \times 3$ unitary matrix $U_R$.
All the matrices in~\eqref{uit5} are $3 \times 3$ symmetric and complex;
therefore,
equation~\eqref{uit5} is in effect a system of 12 real equations for 
the 12 unknowns---$m_{4,5,6}$ and the nine parameters of $U_R$---that we need
for the computation of the $X_{\ell_1 \ell_2}$.
We stress that this parameter counting only serves to demonstrate
the theoretical possibility of obtaining the $X_{\ell_1 \ell_2}$
from equation~\eqref{uit5};
when one attempts to do it numerically,
equation~\eqref{uit5} may sometimes prove difficult or impossible to solve.

In practice,
we reduce the number of parameters by applying assumption~B,
\textit{i.e.}\ the \emph{reality assumption}.
Concretely,
we set
\bs
\label{real}
\ba
& & e^{- i \hat \alpha} = e^{- 2 i \hat \beta} = \bone,
\\
& & e^{i \delta} = - 1\ \mbox{in}\ U_\mathrm{PMNS},
\label{delta}
\\
& & e^{- 2 i \alpha} = 1,
\\
& & d_\ell\ \mbox{real}\ (\ell = e,\mu,\tau),
\\
& & \delta_\ell\ \mbox{real}\ (\ell = e,\mu,\tau).
\ea
\es
In~\eqref{delta} we have opted for $\delta = \pi$,
which is phenomenologically preferred
over $\delta = 0$~\cite{esteban,capozzi,desalas,esteban2020}.
Using the assumptions~\eqref{real},
the symmetric matrix in the left-hand side of~\eqref{uit5} is real,
hence it has just six degrees of freedom.
Then,
the matrix $U_R$ may be written
\be\label{UR1}
U_R = U^\prime_R \times \mathrm{diag} \left( \epsilon_4,\, \epsilon_5,\,
\epsilon_6 \right),
\ee
where $U_R^\prime \in SO(3)$ is a real matrix
parameterized by three angles
and the $\epsilon_{3+j}$ may be either $1$ or $i$.
Equation~\eqref{uit5} is then used
to determine the three angles of $U_R^\prime$
and the three $\epsilon_{3+j}^2\, m_{3+j}$;
the latter are either positive,
if $\epsilon_{3+j} = 1$,
or negative,
if $\epsilon_{3+j} = i$.

Let us take stock of the (real) parameters in the game,
after having performed the simplification stated in the previous paragraph.
From the neutrino oscillation data,
both the two mass-squared differences among the three light-neutrino masses
and the three mixing angles in $U_\mathrm{PMNS}$ are known
and they are used as input.
There are then 15 unknown parameters
in~\eqref{uit5}: 
the lightest neutrino mass,
\textit{viz.}\ $m_1$ for normal ordering
and $m_3$ for inverted ordering of the neutrino masses,
$M_{3,4}$,
$d_\ell$,
$\delta_\ell$,
$\epsilon_{3+j}^2\, m_{3+j}$ for $j = 1, 2, 3$,
and the three angles in $U_R$.
As we shall see in the next section,
there are in addition the three parameters $\gamma_\ell$,
which do not appear
in~\eqref{uit5}
but occur in the BRs.

\section{Decay rates}
\label{branching}

Repeating the result of~\cite{aeikens}, 
the decay rates are given by
\bs
\label{fjfio}
\ba
\Gamma \left( \mu^- \to e^- e^+ e^- \right) &=&
\frac{m_\mu}{6144 \pi^3} \left[ \frac{3}{4} \left( \frac{1}{M_3^4}
  + \frac{1}{M_4^4} \right) + \frac{1}{2 M_3^2 M_4^2} \right]
\no & & \times
\left| X_{\mu e} \right|^2 \left| \gamma_e \right|^2\,
\left( \left| A_{\mu e} \right|^2 + \left| A_{e \mu} \right|^2 \right),
\\
\Gamma \left( \tau^- \to e^- e^+ e^- \right) &=&
\frac{m_\tau}{6144 \pi^3} \left[ \frac{3}{4} \left( \frac{1}{M_3^4}
  + \frac{1}{M_4^4} \right) + \frac{1}{2 M_3^2 M_4^2} \right]
\no & & \times
\left| X_{\tau e} \right|^2 \left| \gamma_e \right|^2\,
\left( \left| A_{\tau e} \right|^2 + \left| A_{e \tau} \right|^2 \right),
\\
\Gamma \left( \tau^- \to e^- \mu^+ \mu^- \right) &=&
\frac{m_\tau}{6144 \pi^3} \left( \frac{1}{M_3^4} + \frac{1}{M_4^4} \right)
\left| X_{\tau e} \right|^2 \left| \gamma_\mu \right|^2\,
\left( \left| A_{\tau e} \right|^2 + \left| A_{e \tau} \right|^2 \right),
\\
\Gamma \left( \tau^- \to \mu^- \mu^+ \mu^- \right) &=&
\frac{m_\tau}{6144 \pi^3} \left[ \frac{3}{4} \left( \frac{1}{M_3^4}
  + \frac{1}{M_4^4} \right) + \frac{1}{2 M_3^2 M_4^2} \right]
\no & & \times
\left| X_{\tau \mu} \right|^2 \left| \gamma_\mu \right|^2\,
\left(\left| A_{\tau \mu} \right|^2 + \left| A_{\mu \tau} \right|^2 \right),
\\
\Gamma \left( \tau^- \to \mu^- e^+ e^- \right) &=&
\frac{m_\tau}{6144 \pi^3} \left( \frac{1}{M_3^4} + \frac{1}{M_4^4} \right)
\left| X_{\tau \mu} \right|^2 \left| \gamma_e \right|^2\,
\left( \left| A_{\tau \mu} \right|^2 + \left| A_{\mu \tau} \right|^2 \right),
\ea
\es
where we have used the approximation that the final-state charged leptons
are massless,
and
\ba
A_{\ell_1 \ell_2} &=&
\frac{\sqrt{2}}{v}
\left( m_{\ell_1}^2 - m_{\ell_2}^2 \right) m_{\ell_1}
\delta_{\ell_1}^\ast d_{\ell_2}
+ m_{\ell_1}^2 \gamma_{\ell_1} \left( \delta_{\ell_1}^\ast \delta_{\ell_2}
- d_{\ell_1}^\ast d_{\ell_2} \right)
\no & &
+ \frac{m_{\ell_1} m_{\ell_2}}{2}\, \gamma_{\ell_2}
\left( 3 d_{\ell_1}^\ast d_{\ell_2}
- \delta_{\ell_1}^\ast \delta_{\ell_2} \right)
- \frac{m_{\ell_2}^2}{2}\, \gamma_{\ell_1} \left( \delta_{\ell_1}^\ast \delta_{\ell_2}
+ d_{\ell_1}^\ast d_{\ell_2} \right)
\no & &
+ \frac{v}{\sqrt{2}}\, m_{\ell_2} \gamma_{\ell_1}
\left( d_{\ell_1}^\ast \gamma_{\ell_2} \delta_{\ell_2}
- \delta_{\ell_1}^\ast \gamma_{\ell_2}^\ast d_{\ell_2} \right)
+ \frac{v}{\sqrt{2}}\, m_{\ell_1}
\left( \delta_{\ell_1}^\ast \left| \gamma_{\ell_2} \right|^2 d_{\ell_2}
- \gamma_{\ell_1}^2 d_{\ell_1}^\ast \delta_{\ell_2} \right).
\label{A}
\ea
We stress that assumption~A is responsible
for the relatively simple form of the decay rates.

\subsection{A prediction}

Taking ratios of decay rates of the $\tau$,
we obtain ratios of BRs.
Defining
\be
x \equiv \left( \frac{M_3}{M_4} \right)^2
\quad \mbox{and} \quad
y \equiv \left| \frac{\gamma_\mu}{\gamma_e} \right|^2,
\ee
we obtain 
\bs
\label{moiop}
\ba
\frac{\mathrm{BR} \left( \tau^- \to e^- \mu^+ \mu^- \right)}{\mathrm{BR}
  \left( \tau^- \to e^- e^+ e^- \right)}
&=& y\ \frac{4 x^2 + 4}{3 x^2 + 3 + 2 x},
\\
\frac{\mathrm{BR} \left( \tau^- \to \mu^- \mu^+ \mu^- \right)}{\mathrm{BR}
  \left( \tau^- \to \mu^- e^+ e^- \right)}
&=& y\ \frac{3 x^2 + 3 + 2 x}{4 x^2 + 4}.
\ea
\es
This implies
\be
\sqrt{\frac{\mathrm{BR} \left( \tau^- \to e^- e^+ e^- \right)\,
    \mathrm{BR} \left( \tau^- \to \mu^- \mu^+ \mu^- \right)}{\mathrm{BR}
    \left( \tau^- \to e^- \mu^+ \mu^- \right)\, \mathrm{BR}
    \left( \tau^- \to \mu^- e^+ e^- \right)}}
= \frac{3 x^2 + 3 + 2 x}{4 x^2 + 4}.
\label{cjfui}
\ee
The maximum of the function
in the right-hand side of~\eqref{cjfui} is~$1$
at $x = 1$;
its minimum is $3/4$ at $x = 0$ and $x = \infty$.
Therefore,
we have the following
\emph{prediction}:
\begin{quote}
The ratio
$\frac{\displaystyle \mathrm{BR} \left( \tau^- \to e^- e^+ e^- \right)\,
  \mathrm{BR} \left( \tau^- \to \mu^- \mu^+ \mu^- \right)}
 {\displaystyle \mathrm{BR}
  \left( \tau^- \to e^- \mu^+ \mu^- \right)\, \mathrm{BR}
  \left( \tau^- \to \mu^- e^+ e^- \right)}$
should lie between ${\displaystyle \frac{9}{16}}$ and $1$.
\end{quote}
This is a non-trivial result of our model,
provided assumption~A holds.

\subsection{Suppressing $\mu^- \to e^- e^+ e^-$}
\label{suppressing}

With the mean lives $\tau_\mu$ and $\tau_\tau$ of muon and tau,
respectively,
it follows from~\eqref{fjfio} that
\be
\label{djsis}
R_\mathrm{BR} \equiv \frac{\mbox{BR} \left( \mu^- \to
  e^- e^+ e^- \right)}{\mbox{BR} \left( \tau^- \to e^- e^+ e^- \right)}
= \frac{\tau_\mu m_\mu}{\tau_\tau m_\tau}\ R_X R_A
= 0.45 \times 10^6 \left( R_X R_A \right),
\ee
where
\be
\label{RXRA}
R_X \equiv \left| \frac{X_{\mu e}}{X_{\tau e}} \right|^2
\quad \mbox{and} \quad
R_A \equiv \frac{\left| A_{\mu e} \right|^2
  + \left| A_{e \mu} \right|^2}{\left| A_{\tau e} \right|^2
  + \left| A_{e \tau} \right|^2}.
\ee

The extant experimental upper bounds on the BRs
are given in table~\ref{LFVlimits}.
\begin{table}[ht]
  \begin{center}
    \renewcommand{\arraystretch}{1.1}
    \begin{tabular}{|rcl|} \hline
      $\mbox{BR} \left( \mu^- \to e^- e^+ e^- \right)$
      &$<$& $1.0 \times 10^{-12}$ \\
      $\mbox{BR} \left( \tau^- \to e^- e^+ e^- \right)$
      &$<$& $2.7 \times 10^{-8}$ \\
      $\mbox{BR} \left( \tau^- \to e^- \mu^+ \mu^- \right)$
      &$<$& $2.7 \times 10^{-8}$ \\
      $\mbox{BR} \left( \tau^- \to \mu^- \mu^+ \mu^- \right)$
      &$<$& $2.1 \times 10^{-8}$ \\
      $\mbox{BR} \left( \tau^- \to \mu^- e^+ e^- \right)$
      &$<$& $1.8 \times 10^{-8}$
      \\ \hline
    \end{tabular}
  \end{center}
  \caption{The experimental upper bounds on the branching ratios.
    The bounds are 90\%~CL and have been taken from~\cite{rpp}.
    \label{LFVlimits}}
\end{table}
In the future,
it is expected that the experimental sensitivity
on $\mbox{BR} \left( \mu^- \to e^- e^+ e^- \right)$
will reach $\sim \! 10^{-16}$~\cite{Mu3e},
while the sensitivity on the BRs of the four LFV $\tau$ decays
may be increased by one order of magnitude to $\sim \! 10^{-9}$
either at a Super $B$ factory~\cite{aushev}
or at the High Luminosity LHC~\cite{cerri,abdul},
and even reach $\sim \! 10^{-10}$ at Belle II~\cite{Belle-II}.

We
will be
interested in obtaining parameter-space points
for which all the BRs are below the extant experimental bounds
but above the expected future sensitivities.
Using the present experimental upper bound $10^{-12}$
on $\mbox{BR} \left( \mu^- \to e^- e^+ e^- \right)$ and taking,
for definiteness,
the future sensitivity on the BRs of the $\tau^-$ decays to be $10^{-9}$,
we obtain from~\eqref{djsis} that
\be
\label{ncvkdp}
R_X R_A \lesssim 2 \times 10^{-9}
\ee
for such points.
This may happen either because $R_X$ is very small,
or $R_A$ is very small,
or both.
Focussing specifically on $R_A$,
by using $m_{\ell_2} \ll m_{\ell_1} \ll v$
together with the assumption that all the Yukawa couplings are real,
we read off from~\eqref{A} the dominant terms
\bs
\ba
A_{\mu e} &\approx& \frac{v m_\mu}{\sqrt{2}}
\left( \gamma_e^2 d_e \delta_\mu - \gamma_\mu^2 d_\mu \delta_e \right),
\\
A_{e \mu} &\approx& \frac{v m_\mu}{\sqrt{2}}\, \gamma_e \gamma_\mu
\left( d_e \delta_\mu - d_\mu \delta_e \right).
\ea
\es
Therefore,
in order to obtain a small $R_A$
both $d_e \delta_\mu - d_\mu \delta_e$ and $\gamma_e^2 - \gamma_\mu^2$
should be small.

\section{Numerical procedure}
\label{numerical}

Solving~\eqref{uit5} means finding $m_4$,
$m_5$,
$m_6$ and the matrix $U_R$. 
The latter is parameterized just as $U_L$ in~\eqref{UL},
\textit{i.e.}\ its elements are given by
\bs
\ba
\left( U_R \right)_{11} &=& C_{12} C_{13}
\exp{\left[ i \left( \alpha^R_1 + \beta^R_1 \right) \right]},
\\
\left( U_R \right)_{12} &=& S_{12} C_{13}
\exp{\left[ i \left( \alpha^R_1 + \beta^R_2 \right) \right]},
\\
\left( U_R \right)_{13} &=& S_{13}
\exp{\left[ i \left( \alpha^R_1 + \beta^R_3 - \delta^R \right) \right]},
\\
\left( U_R \right)_{21} &=&
\left[ - S_{12} C_{23} - C_{12} S_{23} S_{13} \exp{\left( i \delta^R \right)}
  \right] 
\exp{\left[ i \left( \alpha^R_2 + \beta^R_1 \right) \right]},
\\
\left( U_R \right)_{22} &=&
\left[ C_{12} C_{23} - S_{12} S_{23} S_{13} \exp{\left( i \delta^R \right)} \right]
\exp{\left[ i \left( \alpha^R_2 + \beta^R_2 \right) \right]},
\\
\left( U_R \right)_{23} &=& S_{23} C_{13}
\exp{\left[ i \left( \alpha^R_2 + \beta^R_3 \right) \right]},
\\
\left( U_R \right)_{31} &=&
\left[ S_{12} S_{23} - C_{12} C_{23} S_{13} \exp{\left( i \delta^R \right)} \right]
\exp{\left[ i \left( \alpha^R_3 + \beta^R_1 \right) \right]},
\\
\left( U_R \right)_{32} &=&
\left[ - C_{12} S_{23} - S_{12} C_{23} S_{13} \exp{\left( i \delta^R \right)}
  \right] 
\exp{\left[ i \left( \alpha^R_3 + \beta^R_2 \right) \right]},
\\
%%%%% \left( U_R \right)_{23} &=& C_{23} C_{13}
\left( U_R \right)_{33} &=& C_{23} C_{13}
\exp{\left[ i \left( \alpha^R_3 + \beta^R_3 \right) \right]},
\ea
\es
where $S_{ij} = \sin{\theta^R_{ij}}$ and $C_{ij} = \cos{\theta^R_{ij}}$. 
However,
following the reality assumption~\eqref{real},
the matrix in the left-hand side of~\eqref{uit5} is real,
hence $U_R$ is real as well,
apart from possible imaginary factors
$\epsilon_{3+j}$ in~\eqref{UR1}.
In order to avoid finding the same solutions of~\eqref{uit5}
several times in different conventions,
we fix the phases in $U_R$
as $\delta^R = \alpha^R_1 = \alpha^R_2 = \pi$
and $\alpha^R_3 = \beta^R_1 = \beta^R_2 = \beta^R_3 = 0$,\footnote{This
  choice is arbitrary;
  in principle,
  many other phase fixings would be just as good.}
while simultaneously we allow the $\epsilon_{3+j}^2 m_{3+j}$
to be either positive or negative and the angles $\theta_{ij}^R$
to be in any quadrant.

The left-hand side of~\eqref{uit5} is determined in the following way:
choosing normal mass ordering of the light neutrinos,\footnote{We have not
  considered the case of inverted mass ordering,
  which is disfavoured by the phenomenological fits.
  We note,
  however,
  that in a recent analysis~\cite{esteban2020}
  the preference for normal ordering has decreased.} 
the mass $m_1$ is an input and
\be 
m_2 = \sqrt{m_1^2 + \Delta m_{21}^2}
\quad \mbox{and} \quad 
m_3 = \sqrt{m_1^2 + \Delta m_{31}^2}.
\ee
For the mass-squared differences and the mixing angles in $U_\mathrm{PMNS}$ 
we take the best-fit values of~\cite{esteban}:
\be
\label{inputdata}
\begin{array}{l}
\Delta m_{21}^2 = 7.39 \times 10^{-5}\,\mathrm{eV}^2, \quad
\Delta m_{31}^2 = 2.525 \times 10^{-3}\,\mathrm{eV}^2,
\\*[2mm]
\sin^2{\theta_{12}} = 0.310, \quad
\sin^2{\theta_{13}} = 0.02241, \quad
\sin^2{\theta_{23}} = 0.580.
\end{array}
\ee

Our fitting program consists of two parts.
In the first part,
the matrix equation~\eqref{uit5} is solved by using a minimization procedure
wherein the function $\chi_{\mathrm{eq}}^2$,
given in~\eqref{chi2_eq} below,
is adjusted to zero with high precision. 
In this part of the program 
all the parameters that occur in the branching ratios,
except the Yukawa couplings $\gamma_\ell$,
are determined.
In the second part of the program,
we use the parameters obtained in the first part
and we search for $\gamma_\ell$ such that either several
or all five branching ratios
are within the future experimental reach;
this is done with the help of the function $\chi_{\mathrm{br}}^2$
given in~\eqref{chi2_br} below.

The function $\chi_{\mathrm{eq}}^2$ is constructed in the following way.
Let $\left( \mc_\nu^\mathrm{exp} \right)_{ij}$
and $\left( \mc_\nu^\mathrm{theor} \right)_{ij}$
be the matrix elements of the matrices
in the left-hand and right-hand sides,
respectively,
of~\eqref{uit5}.
Then,
the function that we
minimize is\footnote{This function 
  is appropriate
  for both cases of a complex or real $\mc_\nu$;
  in our actual practice,
  we only use it in the real case.}
\be
\label{chi2_eq}
\chi_{\mathrm{eq}}^2 =
\sum_{1 \leq i \leq j \leq 3} \left[ \left( f_{ij}^\mathrm{mod} \right)^2
+ \left( f_{ij}^\mathrm{arg} \right)^2 \right]
\ee
with
\bs
\label{fff}
\ba
f_{ij}^\mathrm{mod} &=& 
\frac{\displaystyle | \left( \mc_\nu^\mathrm{exp} \right)_{ij} | - 
| \left( \mc_\nu^\mathrm{theor} \right)_{ij} |}%
{\displaystyle | \left( \mc_\nu^\mathrm{exp} \right)_{ij} | + 
| \left( \mc_\nu^\mathrm{theor} \right)_{ij} |}\, ,
\\*[1mm]
f_{ij}^\mathrm{arg} &=& 
\left\{
\begin{array}{ccc}
\frac{\displaystyle \arg{\left( \mc_\nu^\mathrm{exp} \right)_{ij}} - 
\arg{\left( \mc_\nu^\mathrm{theor} \right)_{ij}}}
{\displaystyle \arg{\left( \mc_\nu^\mathrm{exp} \right)_{ij}} + 
\arg{\left( \mc_\nu^\mathrm{theor} \right)_{ij}}}
&\Leftarrow& 
\arg{\left( \mc_\nu^\mathrm{exp} \right)_{ij}} \neq 0,
\\*[5mm]
\arg{\left( \mc_\nu^\mathrm{theor} \right)_{ij}}
&\Leftarrow& 
\arg{\left( \mc_\nu^\mathrm{exp} \right)_{ij}} = 0.
\end{array} 
\right.  
\ea
\es

In the first part of our fitting program we proceed in the following way.
The mass-squared differences $\Delta m^2_{21}$ and $\Delta m^2_{31}$
and the lepton mixing angles $\theta_{12}$,
$\theta_{13}$,
and $\theta_{23}$ are fixed to their best-fit values~\cite{esteban}. 
In section~\ref{light} we have already stated the values of $v$,
$m_Z$,
and $m_H$ used in our code.
Nine parameters---the masses $M_3$ and $M_4$ of the new scalars, 
the mass $m_1$ of the lightest neutrino,
and the real Yukawa couplings $d_\ell$ and $\delta_\ell$
for $\ell = e, \mu, \tau$---are inputted into~\eqref{uit5};
that matrix equation is solved by minimizing $\chi_{\mathrm{eq}}^2$
with respect to the six parameters $\theta^R_{ij}$
and $\epsilon_{3+j}^2\, m_{3+j}$,
which form the output of~\eqref{uit5}.
We consider~\eqref{uit5} to be solved when $\chi_{\mathrm{eq}}^2 < 10^{-16}$;
the resulting set of 15 parameters is then saved 
for usage in the second part of the fitting program.

Note that,
since we use a minimization procedure, 
we may as well explore the full parameter space
and minimize $\chi_{\mathrm{eq}}^2$
with respect to all 15 parameters simultaneously.
It is also possible to choose
any subspace in the 15-dimensional parameter space
and to perform the minimization of $\chi_{\mathrm{eq}}^2$
in that subspace;
indeed,
in the following we shall do precisely this,
by either fixing or imposing restrictions on the ranges
of some of the input parameters prior to minimization
of $\chi_{\mathrm{eq}}^2$.

The function $\chi_{\mathrm{br}}^2$ is constructed in the following way:
\ba
\chi_{\mathrm{br}}^2
&=& 
\sum_{i=1}^5
\left[
\Theta \left( \mbox{BR}_i^\mathrm{bound} - \mbox{BR}_i^\mathrm{theor} \right) 
\left( \frac{\mbox{BR}_i^\mathrm{bound}}{\mbox{BR}_i^\mathrm{theor}} \right)^2 
\right. \nonumber \\ &&
\left. + 
\Theta \left(\mbox{BR}_i^\mathrm{theor} - \mbox{BR}_i^\mathrm{bound} \right)
\left( \frac{\mbox{BR}_i^\mathrm{theor} - \mbox{BR}_i^\mathrm{bound}}{k} \right)^2 
\,\right],
\label{chi2_br}
\ea
where the index $i$ runs over the five BRs,
$\Theta$ is the Heaviside step function, 
$\mbox{BR}_i^\mathrm{bound}$ denotes the experimental upper bound on each BR
(these are the bounds given in table~\ref{LFVlimits}),
$\mbox{BR}_i^\mathrm{theor}$ is the calculated value of the BR, 
and $k$ is a small number that is meant
to give a kick to the minimization algorithm
whenever the calculated value is larger than the experimental upper bound.
Note that the minimum possible value of $\chi_{\mathrm{br}}^2$ is five,
which materializes in the limit where all five calculated BRs
are just a little smaller than the experimental bound on 
the corresponding BR.\footnote{One might object that 
by minimizing the function~\eqref{chi2_br}
one would almost always end up with points
having all the computed $\mathrm{BR}_i^\mathrm{theor}$
very close to their respective present experimental upper bounds
$\mathrm{BR}_i^\mathrm{bound}$.
This does not happen,
though,
because it is quite difficult to reach the minimum value $5$
of $\chi^2_\mathrm{br}$ by just varying
the three parameters $\gamma_{e,\mu,\tau}$.
Actually,
as one can check for instance by looking at figure~\ref{fig4} below,
even after minimizing $\chi^2_\mathrm{br}$ we obtain lots of points with
$\mathrm{BR}_i^\mathrm{theor} \ll \mathrm{BR}_i^\mathrm{bound}$
for some of the five decays.}
The minimization function~\eqref{chi2_br} can handle even situations 
when the calculated BRs and the upper experimental bounds 
differ by many orders of magnitude.
The function $\chi_{\mathrm{br}}^2$ is 
minimized only with respect to the Yukawa couplings $\gamma_\ell$,
because the other parameters have been
fixed already in the first part of the fitting program.
We stress that,
in contrast to $\chi_{\mathrm{eq}}^2$,
it is not necessary to minimize $\chi_{\mathrm{br}}^2$ with high precision,
since our objective is to obtain BRs that are \emph{below}
but not necessarily close to their respective experimental bounds.
We use $10^{-16}$ and $10^{-9}$ as the future experimental sensitivities
on $\mbox{BR} \left( \mu^- \to e^- e^+ e^- \right)$
and the BRs of the $\tau^-$ decays,
respectively.

Often,
we want to compare the results of equation~\eqref{uit5}
with the ones of its tree-level counterpart
\be
\label{uit6}
e^{- i \hat \alpha}\, U_\mathrm{PMNS}^\ast
\left( \hat m\, e^{- 2 i \hat \beta} \right)
U_\mathrm{PMNS}^\dagger\, e^{- i \hat \alpha} =
- \frac{v^2}{2}\, \Delta_1\, U_R^\ast\,
\frac{1}{\widetilde{m}}\, U_R^\dagger\,
\Delta_1.
\ee
Whenever we perform such a comparison,
we use the superscript ``(loop)''
on quantities that arise from the solution of~\eqref{uit5}
and the superscript ``(tree)''
on quantities that arise from the solution of~\eqref{uit6}.
It is one objective of this letter to show that the quantities
with superscript ``(loop)''
may be substantially different from the corresponding quantities
with superscript ``(tree)''.

\section{Results} \label{results}

\subsection{Evolution of the $X_{\ell_1 \ell_2}$}

In this subsection we give two examples of the way
the quantities $X_{\ell_1 \ell_2}$ may change when the input parameters 
are varied.

\begin{figure}[ht]
  \begin{center}
    \includegraphics[width=1.0\textwidth]{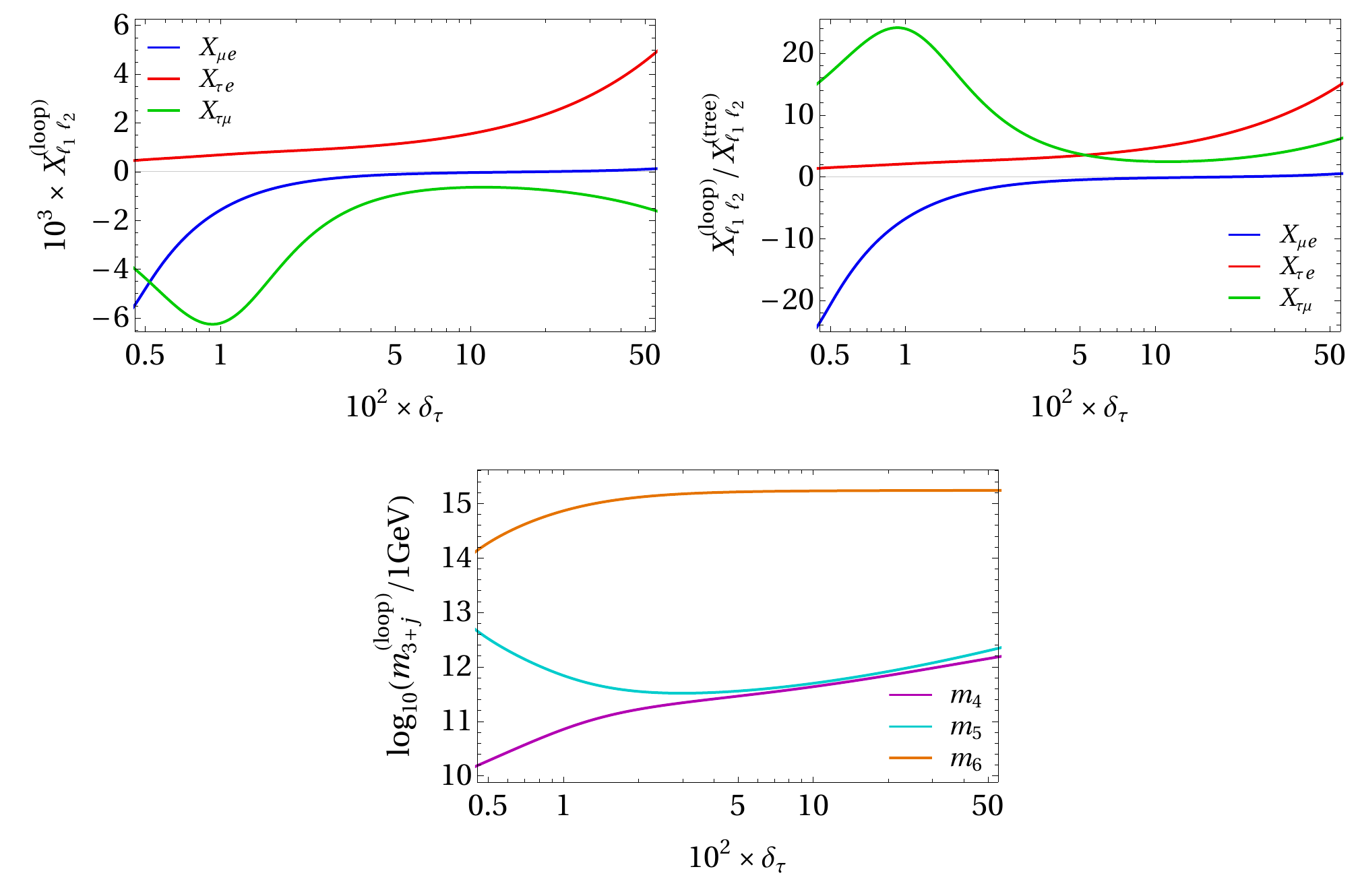}
  \end{center}
  \caption{$X^\mathrm{(loop)}_{\ell_1 \ell_2}$
    (top-left panel),
    $X^\mathrm{(loop)}_{\ell_1 \ell_2} /\, X^\mathrm{(tree)}_{\ell_1 \ell_2}$
    (top-right panel),
    and $m^\mathrm{(loop)}_{4,5,6}$
    (bottom panel)
    against $\delta_\tau$ in a case where
    all other input parameters are kept fixed
    at values given in the main text.
    %%%%% NEW PERIOD
    The definition of the quantities $X_{\ell_1 \ell_2}$ is given in~\eqref{X};
    the definition of the right-handed neutrino masses $m_{4,5,6}$
    is in~\eqref{UR};
    the Yukawa coupling $\delta_\tau$ is defined in~\eqref{Delta1Delta2}.
    \label{fig1}}
\end{figure}

In our first example we fix eight inputs as follows:
$m_1 = 30$\,meV,
$M_3 = 1.5$\,TeV,
$M_4 = 1.6$\,TeV,\footnote{$M_3$ and $M_4$ must be rather close to each other,
  because their difference comes from a coupling in the scalar potential
  that is bounded by unitarity.
  See appendix~A for details.}
$d_e = 0.01$,
$d_\mu = 0.1$,
$d_\tau = 0.001$,
$\delta_e = 1$,
and $\delta_\mu = 0.001$.
We vary $\delta_\tau$ from 0.005 to 0.5 and compute $X_{\mu e}$,
$X_{\tau e}$,
and $X_{\tau \mu}$ for each value of $\delta_\tau$.
In this case $\Delta_1$ is kept fixed,
hence the solution of~\eqref{uit6} is always the same
and produces $X^\mathrm{(tree)}_{\mu e} = 0.000232$,
$X^\mathrm{(tree)}_{\tau e} = 0.000328$,
$X^\mathrm{(tree)}_{\tau \mu} = -0.000260$,
and heavy-neutrino masses $m^\mathrm{(tree)}_4 = 7.18 \times 10^8$\,GeV,
$m^\mathrm{(tree)}_5 = 9.83 \times 10^{10}$\,GeV,
and $m^\mathrm{(tree)}_6 = 7.16 \times 10^{12}$\,GeV.
In figure~\ref{fig1} we display the corresponding quantities
with ``(loop)'' superscript.
In particular,
one observes in the top-left panel of that figure that $X_{\mu e}^\mathrm{(loop)}$
is zero for $\delta_\tau \simeq 0.1$.
In the top-right panel of figure~\ref{fig1} one sees that
both $X^\mathrm{(loop)}_{\mu e}$ and $X^\mathrm{(loop)}_{\tau \mu}$
are one order of magnitude larger than the corresponding tree-level
quantities when $\delta_\tau \lesssim 0.01$,
and the same happens for $X_{\tau e}$ when $\delta_\tau \gtrsim 0.2$.
In the bottom panel of figure~\ref{fig1} one sees that
all three heavy neutrinos are heavier when their masses are computed
by taking into account the loop corrections;
for instance,
$m_6^\mathrm{(loop)} \sim 10^{15}$\,GeV
while $m_6^\mathrm{(tree)} < 10^{13}$\,GeV.

\begin{figure}[ht]
  \begin{center}
   \includegraphics[width=1.0\textwidth]{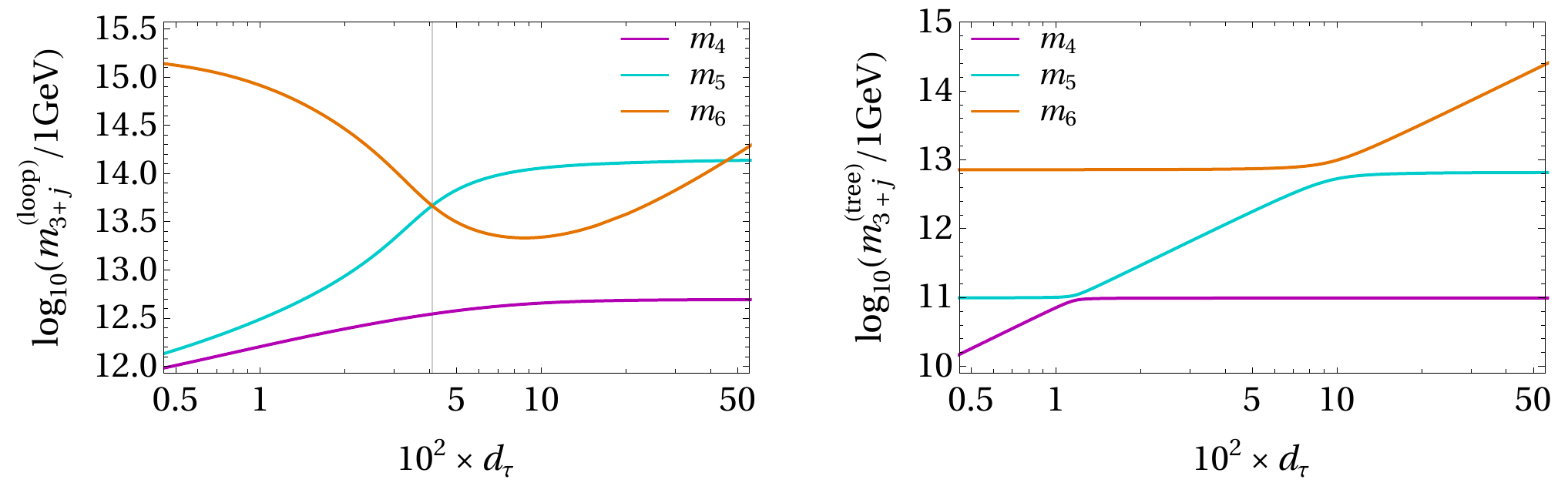}
  \end{center}
  \caption{$m^\mathrm{(loop)}_{4,5,6}$ (left panel)
    and $m^\mathrm{(tree)}_{4,5,6}$ (right panel)
    against $d_\tau$ in a case where all other input parameters
    are kept fixed at values given in the main text.
    %%%%% NEW PERIOD
    The Yukawa coupling $d_\tau$ is defined in~\eqref{Delta1Delta2}.
    \label{fig2}}
\end{figure}

\begin{figure}[ht]
  \begin{center}
    \includegraphics[width=1.0\textwidth]{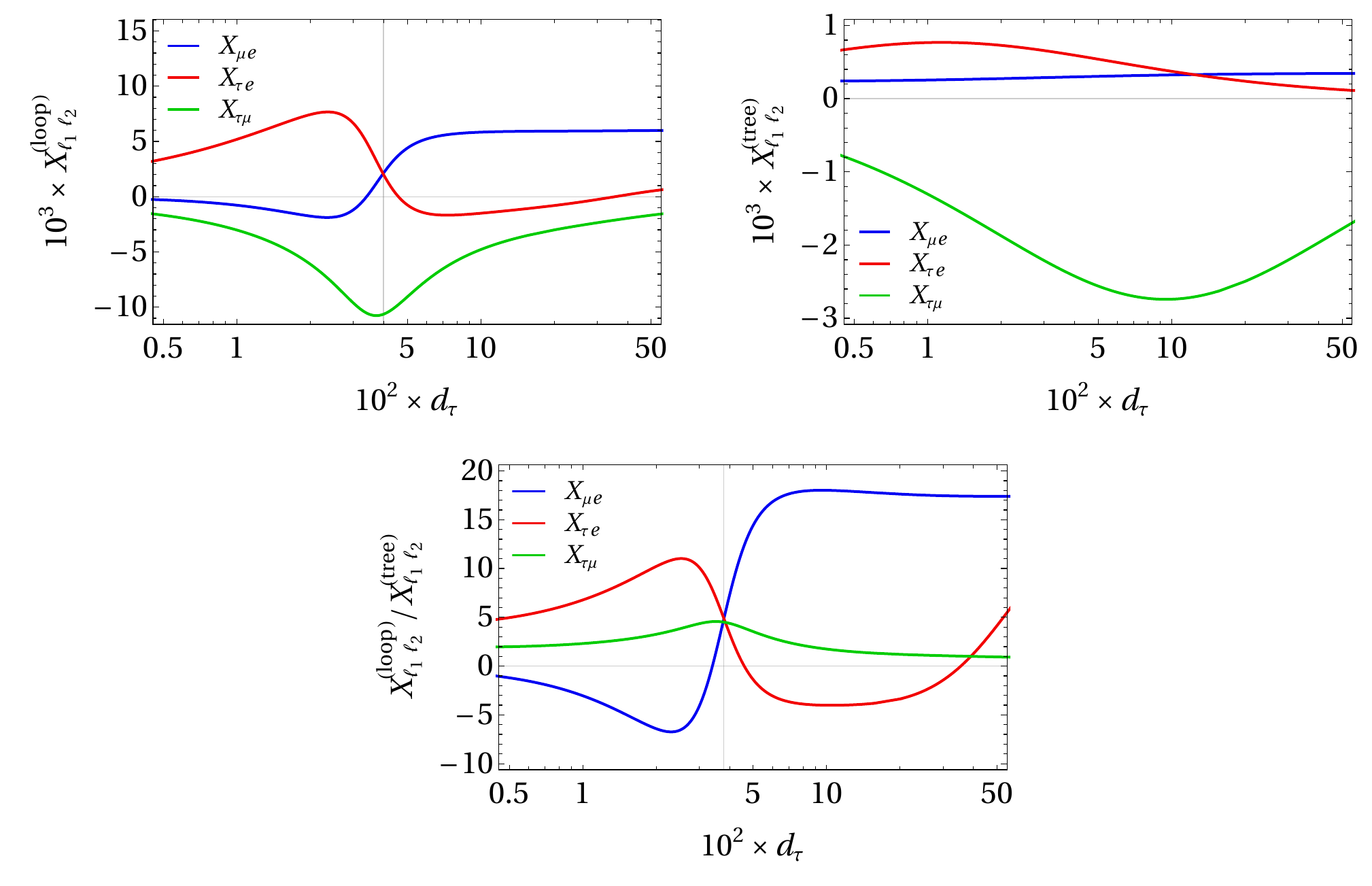}
  \end{center}
  \caption{$X^\mathrm{(loop)}_{\ell_1 \ell_2}$
    (top-left panel),
    $X^\mathrm{(tree)}_{\ell_1 \ell_2}$
    (top-right panel),
    and
    $X^\mathrm{(loop)}_{\ell_1 \ell_2} /\, X^\mathrm{(tree)}_{\ell_1 \ell_2}$
    (bottom panel)
    against $d_\tau$, with the same input as in figure~\ref{fig2}.
    \label{fig3}}
\end{figure}

In our second example we fix eight input parameters as follows:
$m_1 = 30$\,meV,
$M_3 = 1.5$\,TeV,
$M_4 = 1.6$\,TeV,
$d_e = 0.01$,
$d_\mu = 0.1$,
$\delta_e = 1$,
$\delta_\mu = 0.001$,
and $\delta_\tau = 0.1$.
We vary $d_\tau$ from 0.005 to 0.5
and we solve both~\eqref{uit5} and~\eqref{uit6} for each value of $d_\tau$.
The results obtained for the heavy-neutrino masses $m_{3+j}$
are depicted in figure~\ref{fig2},
at the loop level in the left panel and at tree level in the right one.
One sees that,
in the tree-level solution,
the heavy-neutrino masses have a very simple behaviour:
for low $d_\tau$,
$m_4^\mathrm{(tree)}$ increases with $d_\tau$ while
$m_5^\mathrm{(tree)}$ and $m_6^\mathrm{(tree)}$ remain almost constant
(in reality,
they also increase but very slowly);
then,
for intermediate $d_\tau$,
it is $m_5^\mathrm{(tree)}$ that increases at a regular pace while
$m_4^\mathrm{(tree)}$ and $m_6^\mathrm{(tree)}$ remain constant;
finally,
for high $d_\tau$,
$m_6^\mathrm{(tree)}$ increases but
$m_4^\mathrm{(tree)}$ and $m_5^\mathrm{(tree)}$ are stable.
Including the radiative corrections
(left panel of figure~\ref{fig2})
the whole picture changes;
all three heavy-neutrino masses become
one or two orders of magnitude larger,
and moreover $m_5^\mathrm{(loop)}$ and $m_6^\mathrm{(loop)}$
exhibit a peculiar behaviour,
interchanging positions at $d_\tau \approx 0.04$
and then again at $d_\tau \approx 0.45$.
This peculiar behaviour of $m_{5,6}^\mathrm{(loop)}$
has a counterpart in the behaviour of the $X^\mathrm{(loop)}_{\ell_1 \ell_2}$ 
depicted in the top-left panel of figure~\ref{fig3}.
One sees that both $X^\mathrm{(loop)}_{\mu e}$ and $X^\mathrm{(loop)}_{\tau e}$
experience sudden changes close to the point where
%%%%% $m_5^\mathrm{(tree)}$ and $m_6^\mathrm{(tree)}$
%%%%% first interchange positions.
$m_5^\mathrm{(loop)}$ and $m_6^\mathrm{(loop)}$ first interchange positions.
One moreover sees that $X^\mathrm{(loop)}_{\tau e}$ is zero
for two different values of $d_\tau$,
while $X^\mathrm{(loop)}_{\mu e}$ is zero only once.
In the bottom panel of figure~\ref{fig3}
one sees that the
$X^\mathrm{(loop)}_{\ell_1 \ell_2} /\, X^\mathrm{(tree)}_{\ell_1 \ell_2}$
are typically of order 10,
but both $X^\mathrm{(loop)}_{\tau e} /\, X^\mathrm{(tree)}_{\tau e}$
and $X^\mathrm{(loop)}_{\mu e} /\, X^\mathrm{(tree)}_{\mu e}$ have zeros.
It is amusing to note that all three
$X^\mathrm{(loop)}_{\ell_1 \ell_2} /\, X^\mathrm{(tree)}_{\ell_1 \ell_2}$
have approximately the same value at the first point where
$m_5^\mathrm{(loop)}$ and $m_6^\mathrm{(loop)}$ cross.

\subsection{Scatter plots of BRs}

\begin{figure}[t]
  \begin{center}
    \includegraphics[width=1.0\textwidth]{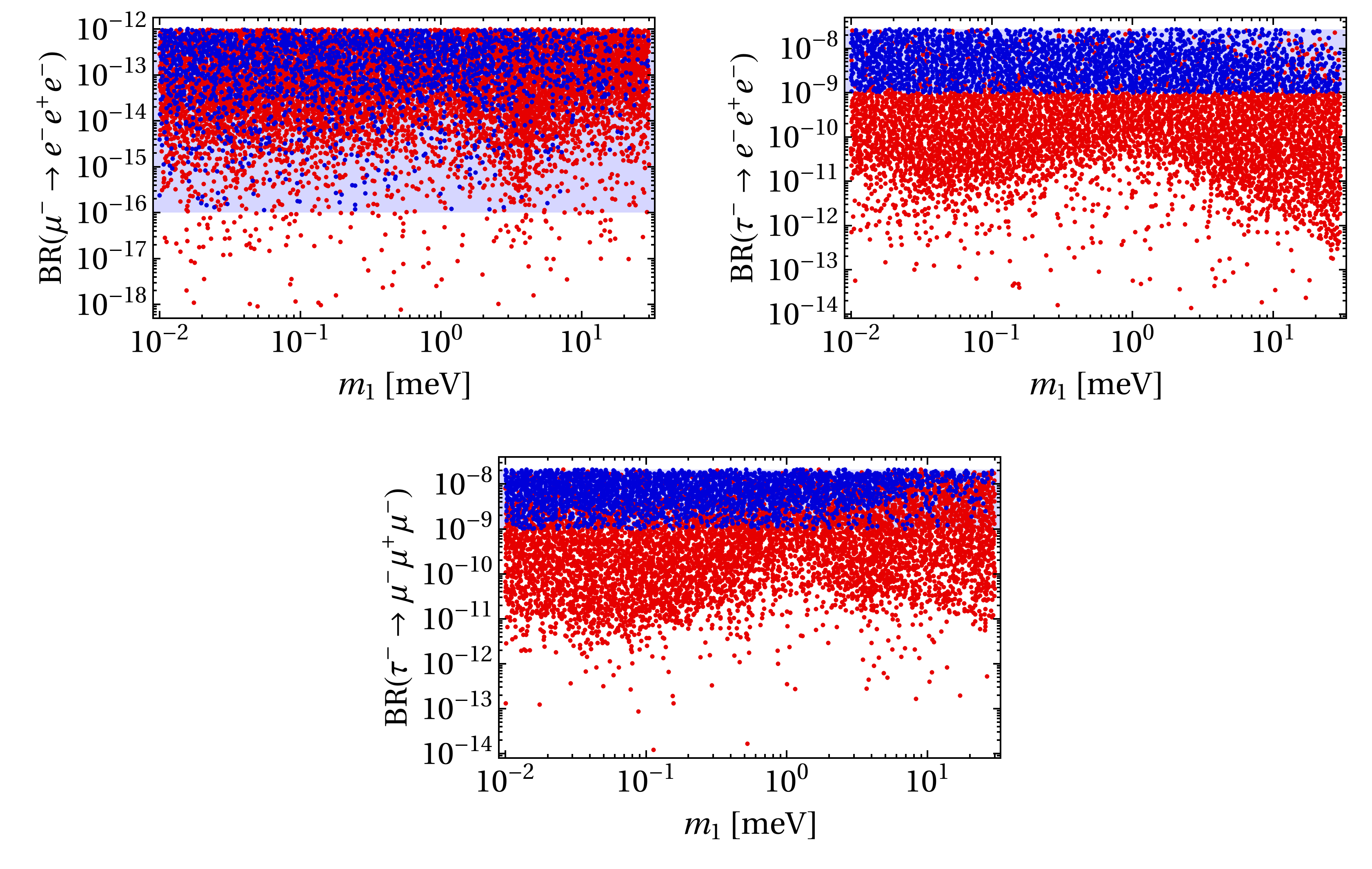}
  \end{center}
  \caption{Scatter plots of
    $\mbox{BR} \left( \mu^- \to e^- e^+ e^- \right)$ (top-left panel),
    $\mbox{BR} \left( \tau^- \to e^- e^+ e^- \right)$ (top-right panel),
    and $\mbox{BR} \left( \tau^- \to \mu^- \mu^+ \mu^- \right)$ (bottom panel)
%%%%%    as functions of $m_1$;
    as functions of the lightest-neutrino mass $m_1$;
    the other inputs are given in~\eqref{nqpeee} and~\eqref{mytiytpqq}.
    In all the displayed points,
    all five BRs satisfy the present experimental bounds
    given in table~\ref{LFVlimits}.
    Blue points have all five BRs larger than 
    the expected future sensitivities,
    while red points allow one or more
    (but not all)
    BRs to be below the future sensitivities.
    The shadowed bands show the ranges
    between the present experimental bounds
    and the future experimental sensitivities,
    \textit{viz.}\ $10^{-16}$
    for $\mbox{BR} \left( \mu^- \to e^- e^+ e^- \right)$
    and $10^{-9}$ for the BRs of the $\tau^-$ decays.
    \label{fig4}}
\end{figure}

In figure~\ref{fig4}
we present scatter plots of the branching ratios
of\footnote{As seen in~\eqref{moiop},
  in our model the branching ratio of $\tau^- \to e^- \mu^+ \mu^-$
  is closely related to the one of $\tau^- \to e^- e^+ e^-$,
  and the branching ratio of $\tau^- \to \mu^- e^+ e^-$
  is related to the one of $\tau^- \to \mu^- \mu^+ \mu^-$.
  For this reason and in order to save space,
  we omit figures for $\mbox{BR} \left( \tau^- \to e^- \mu^+ \mu^- \right)$
  and for $\mbox{BR} \left( \tau^- \to \mu^- e^+ e^- \right)$.}
$\mu^- \to e^- e^+ e^-$,
$\tau^- \to e^- e^+ e^-$,
and $\tau^- \to \mu^- \mu^+ \mu^-$ as functions of $m_1$.
Here as elsewhere in this section we always assume,
for the sake of simplicity,
the neutrino mass ordering to be normal.
To produce the scatter plots, 
%%%%% we chose $m_1$ at random in betweeen $10^{-2}$\,meV to 30~meV,
we chose $m_1$ at random in betweeen $10^{-2}$\,meV to 30\,meV,
prior to the minimization of $\chi^2_\mathrm{eq}$;
larger values of $m_1$ would violate the {\it Planck}~2018 cosmological bound
on the sum of the light-neutrino masses~\cite{Ade:2013zuv}.
Then the BRs are computed,
as described in section~\ref{numerical},
by consecutive minimization of $\chi^2_\mathrm{eq}$ and $\chi^2_\mathrm{br}$ 
with respect to the remaining parameters. 
We restrict the parameter space by adopting the boundary conditions
\bs
\label{nqpeee}
\ba
\label{msjww}
& & 750\,\mathrm{GeV} < M_{3,4} < 2\,\mbox{TeV},
\\*[1mm]
\label{slpwo}
& & M_3^2 - \frac{8 \pi}{3}\, v^2 < M_4^2 < M_3^2 + \frac{8 \pi}{3}\, v^2,
\\*[1mm] & &
%%%%% NEW LABEL
\label{vmdlfo0}
0.05 \leq \left| d_\ell \right|,\, \left| \delta_\ell \right|,\,
\left| \gamma_\ell \right| \leq 0.5 \quad
(\ell=e,\mu,\tau),
\ea
\es
and
\be
\label{mytiytpqq}
10^{11}\,\mbox{GeV} \le m_{4,5,6} \leq 10^{16}\,\mbox{GeV},
\ee
with the $\epsilon_{3+j}^2\, m_{3+j}$ being either positive or negative.
%%%%% NEW PERIOD
Notice that the range of Yukawa couplings
that we have considered in~\eqref{vmdlfo0} is quite restricted
compared to the Yukawa couplings of the charged fermions,
that are known to vary from $\sim 10^{-6}$ to $\sim 1$.

It is worth making a number of comments concerning~\eqref{nqpeee}
and~\eqref{mytiytpqq}:
\begin{enumerate}
\item The lower bound on $M_3$ and $M_4$
  that we have assumed in~\eqref{msjww}
  agrees roughly with the results of a recent analysis~\cite{analysis}
  of 2HDMs furnished with an additional $\mathbbm{Z}_2$ symmetry.
\item In~\eqref{slpwo} the bounds on $M_4$ have been chosen
  in such a way that all the relevant conditions
  on the 2HDM scalar potential are met.
  Namely,
  the difference between $M_3^2$ and $M_4^2$ originates in a coupling
  of the scalar potential that is bounded by unitarity,
  and therefore $\left| M_3^2 - M_4^2 \right|$ cannot be too large.
  See appendix~A for details.
\item Sometimes the solution of~\eqref{uit5}
  requires one of the $m_{3+j}$ to be very large,
  even divergent.
  This is not surprising because,
  when \textit{e.g.}\ $m_6 \to \infty$,
  the contribution of $m_6$ to~\eqref{uit5} simply vanishes.
  Unfortunately,
  though,
  when $m_6 \to \infty$ the $X_{\ell_1 \ell_2}$ diverge.
  We avoid this problem by discarding,
  through the upper bound in~\eqref{mytiytpqq},
  those points where the solution of~\eqref{uit5} requires very large $m_{3+j}$.
\item We have obtained points with values
  of the heavy-neutrino masses as low as $10^9$\,GeV.
  However,
  those points have very low BRs for the decays of the $\tau^-$,
  of order $10^{-12}$.
  In~\eqref{mytiytpqq} we have discarded those points
  by enforcing a lower bound on the heavy-neutrino masses.
\end{enumerate}

\begin{figure}[ht]
  \begin{center}
    \includegraphics[width=1.0\textwidth]{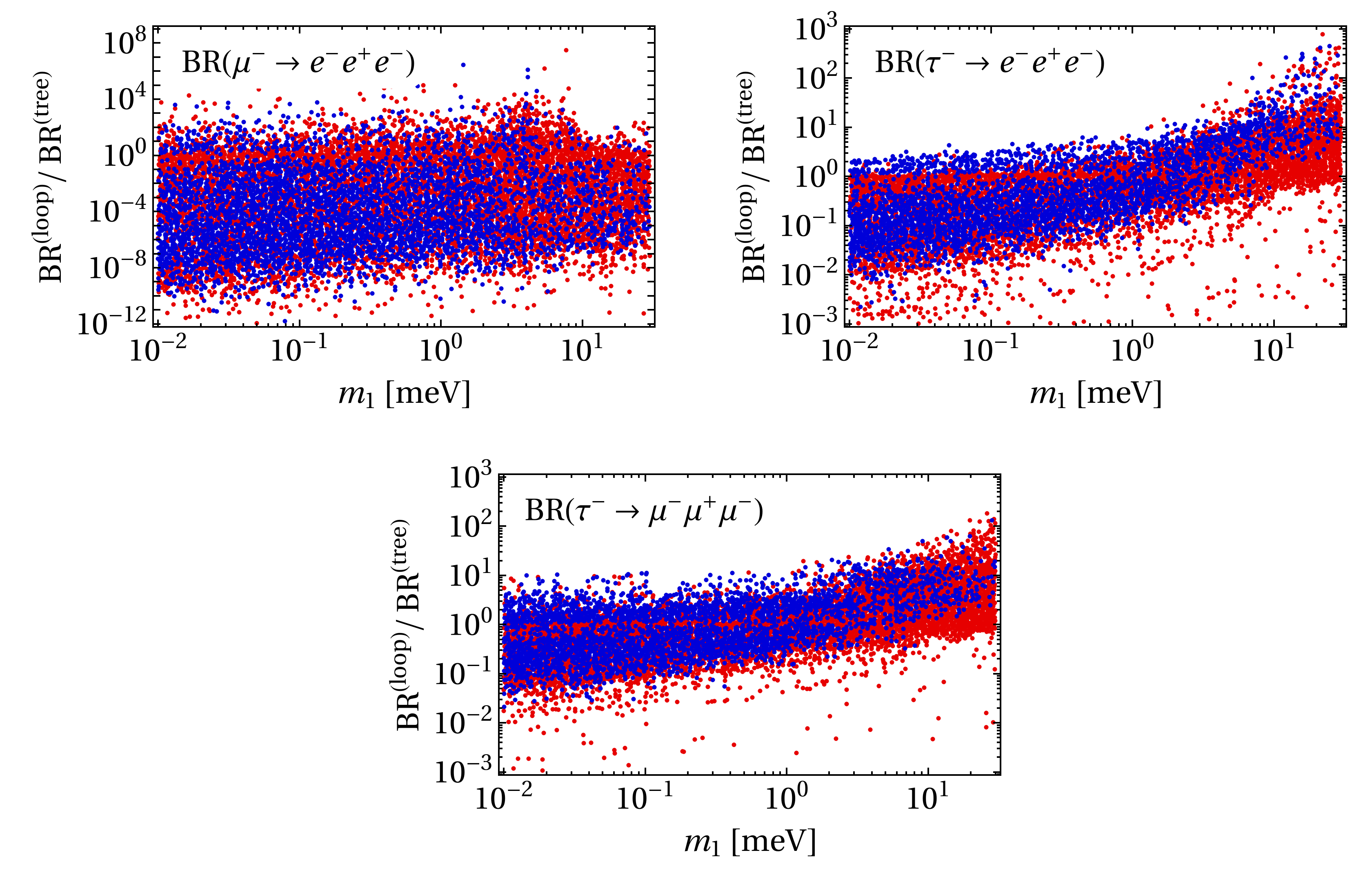}
  \end{center}
  \caption{Scatter plots of
    $\mbox{BR}^\mathrm{(loop)} /\, \mbox{BR}^\mathrm{(tree)}$
    for the decays $\mu^- \to e^- e^+ e^-$ (top-left panel),
    $\tau^- \to e^- e^+ e^-$ (top-right panel),
    and $\tau^- \to \mu^- \mu^+ \mu^-$ (bottom panel).
    The points are the ones used in figure~\ref{fig4},
    with the same colour coding as there.
    \label{fig5}}
\end{figure}

In figure~\ref{fig5} we display the ratios
$\mbox{BR}^\mathrm{(loop)} \left/ \mbox{BR}^\mathrm{(tree)} \right.$
for the same points as in figure~\ref{fig4} and with the same colour notation.
One sees that for the $\tau^-$ decays 
the BRs derived from~\eqref{uit5}
may easily be one or two orders of magnitude
either above or below the corresponding BRs derived from~\eqref{uit6}.
For the decay $\mu^- \to e^- e^+ e^-$ things may be much more dramatic,
with differences of several orders of magnitude;
this happens because either $X_{\mu e}^\mathrm{(loop)}$
or $X_{\mu e}^\mathrm{(tree)}$ frequently become zero.
It is worth mentioning that by allowing for wider ranges of the Yukawa couplings
(for example,
allowing $\left| d_\ell \right|$,
$\left| \delta_\ell \right|$,
and $\left| \gamma_\ell \right|$ to be between 0.001 and 1)
would lead to the ratios $\mbox{BR}^\mathrm{(loop)}
\left/\, \mbox{BR}^\mathrm{(tree)} \right.$ being sometimes much larger;
those ratios could be two orders of magnitude larger or smaller
than is shown in figure~\ref{fig5}.

\subsection{The suppression of $\mu^- \to e^- e^+ e^-$}

\begin{figure}[ht]
  \begin{center}
    \includegraphics[width=1.0\textwidth]{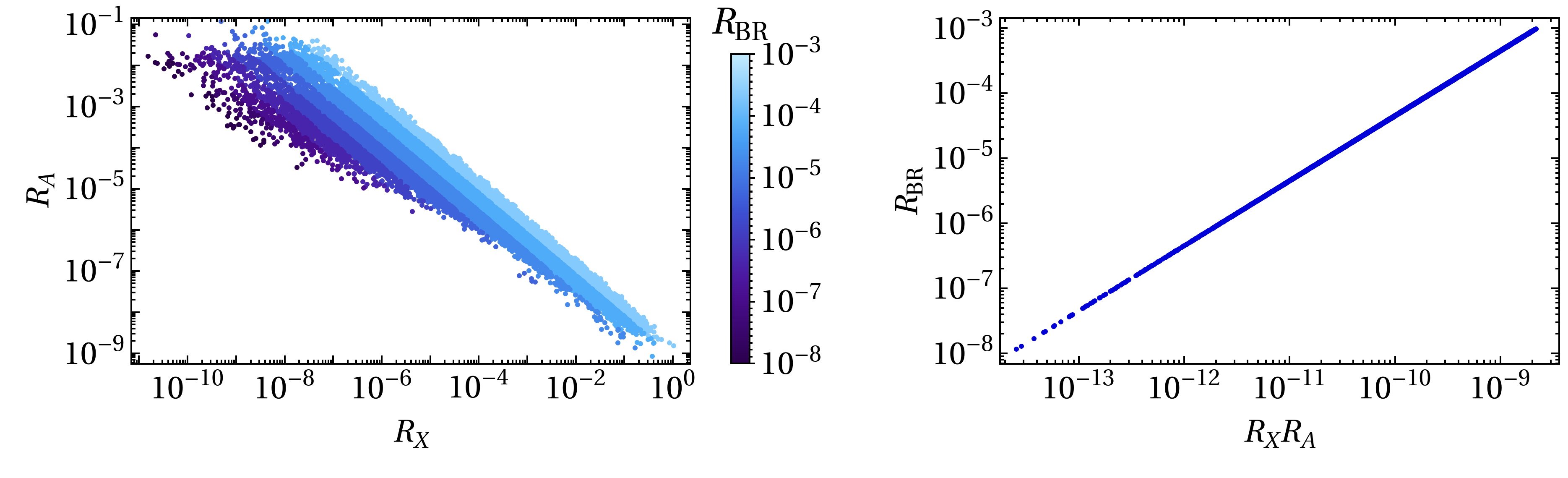}
  \end{center}
  \caption{Scatter plots of $R_A$ \textit{vs.}\ $R_X$
    (left panel)
    and of their product $R_X R_A$ \textit{vs.}\ $R_\mathrm{BR}$
    (right panel)
    for the blue points of figure~\ref{fig4}.
    The definitions of $R_A$,
    $R_X$,
    and $R_\mathrm{BR}$ are given in~\eqref{djsis} and~\eqref{RXRA}.
    \label{fig6}}
\end{figure}

In figure~\ref{fig6} we reuse the blue points of the previous
figures~\ref{fig4} and~\ref{fig5},
\textit{viz.}\ points for which all five BRs are in between
the respective present  upper bounds and future expected sensitivities.
For those points,
we display $R_\mathrm{BR}$ defined in~\eqref{djsis},
and $R_X$ and $R_A$ defined in~\eqref{RXRA}.
In the right panel one sees that the inequality~\eqref{ncvkdp} holds
and that $R_X R_A$ is proportional to $R_\mathrm{BR}$ as stated in~\eqref{djsis}.
In the left panel one sees that the smallness of $R_X R_A$
most of the time occurs because both $R_A$ and $R_X$ are small,
but there is a non-negligible fraction of points where one of them
is extremely small and the other one is not small.

\begin{figure}[ht]
  \begin{center}
    \includegraphics[width=1.0\textwidth]{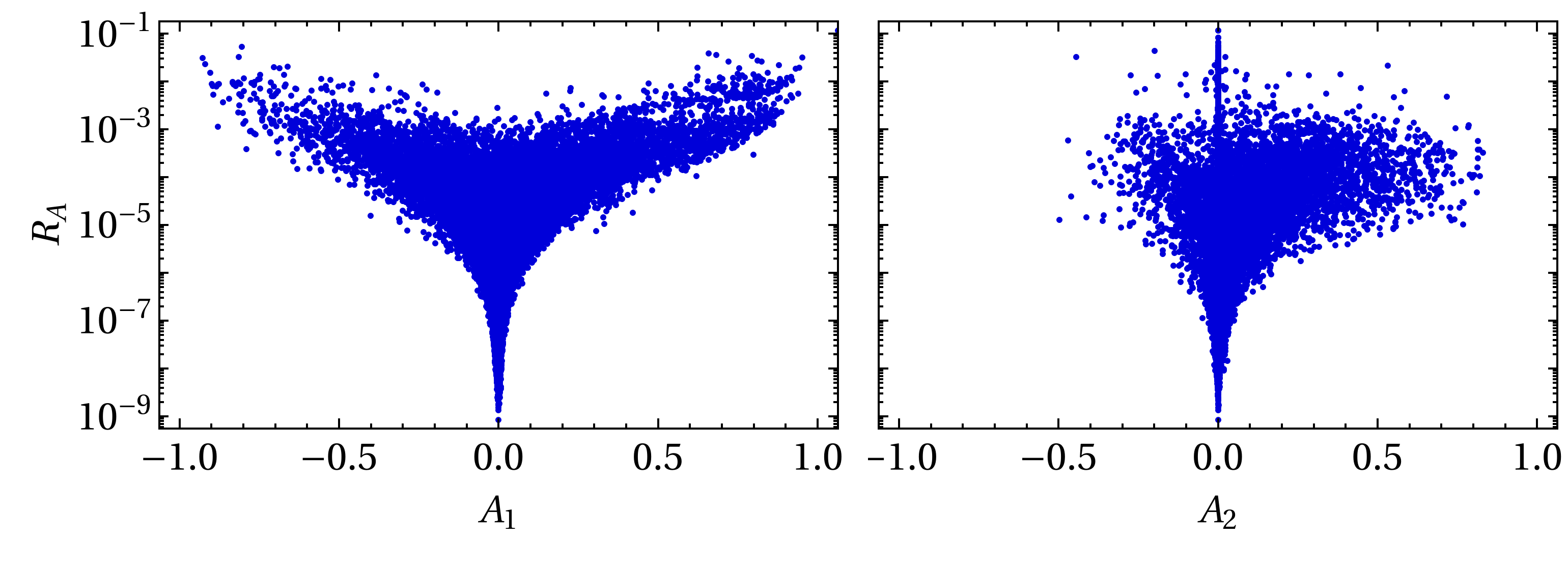}
  \end{center}
  \caption{Scatter plots of $A_1$
    (left panel)
    and $A_2$
    (right panel)
    against $R_A$ for the blue points of figure~\ref{fig4}.
    The definitions of $A_1$ and $A_2$ are given in~\eqref{A1A2}.
    \label{fig7}}
\end{figure}

The discussion at the end of section~\ref{suppressing} suggests that
the smallness of $R_A$ is correlated with the smallness of the asymmetries 
\be
\label{A1A2}
A_1 \equiv \frac{d_e \delta_\mu - d_\mu \delta_e}{d_e \delta_\mu
  + d_\mu \delta_e}
\quad \mbox{and} \quad
A_2 \equiv \frac{\gamma_e^2 - \gamma_\mu^2}{\gamma_e^2 + \gamma_\mu^2}.
\ee
Using the same points as in figure~\ref{fig6},
these asymmetries are displayed in figure~\ref{fig7}.
One sees that
$A_1$ and $A_2$ are indeed very small when $R_A \lesssim 10^{-7}$,
but they may be largish for $R_A$ above that value;
we remind the reader that,
like we saw in figure~\ref{fig6},
the smallness of $R_X R_A$ is often due to the smallness of $R_X$
and not to the smallness of $R_A$,
or vice-versa.

\subsection{Benchmark points}

In table~\ref{bench} we produce three benchmark points.
\begin{table}[ht]
  \begin{center}
    \begin{tabular}{|c|c|c|c|} \hline
       & Point~1 & Point~2 & Point~3 \\ \hline
      $d_e$ & $-0.1007921873$ & $-0.4760159332$ & $-0.1486369437$ \\
      $d_\mu$ & $-0.1008806975$ & $-0.3515469881$ & $-0.1350920928$ \\
      $d_\tau$ & $0.4284126498$ & $-0.1867255478$ & $0.4815723378$ \\
      $\delta_e$ & $0.2866985699$ & $0.06004123429$ & $-0.4967063119$ \\
      $\delta_\mu$ & $0.2867857061$ & $-0.4838389436$ & $0.1463403266$ \\
      $\delta_\tau$ & $-0.05682546538$ & $-0.4996927131$ & $-0.4386169548$ \\
      $m_1$\,(meV) & $19.97920246$ & $21.24771538$ & $0.01193047926$ \\
      $M_3$\,(GeV) & $1\,850.763353$ & $1\,687.165806$ & $948.4168772$ \\
      $M_4$\,(GeV) & $1\,907.962751$ & $1\,753.477583$ & $822.8728412$ \\ \hline
      $\epsilon_4^2 m_4$\,(GeV) & $-1.431876108 \times 10^{14}$ &
      $-7.187027731 \times 10^{15}$ & $8.630665168 \times 10^{13}$ \\
      $\epsilon_5^2 m_5$\,(GeV) & $-4.522247054 \times 10^{13}$
      & $-1.190583685 \times 10^{14}$
      & $6.484748230 \times 10^{12}$ \\
      $\epsilon_6^2 m_6$\,(GeV) & $9.182836790 \times 10^{15}$
      & $-3.431144882 \times 10^{14}$
      & $2.627559538 \times 10^{15}$ \\
      $\theta^R_{12}$\,(rad) & $2.974230185$ & $3.940376049$ & $0.2251793380$ \\
      $\theta^R_{13}$\,(rad) & $3.322622001$ & $2.779495689$ & $1.533482136$ \\
      $\theta^R_{23}$\,(rad) & $2.520334568$ & $1.410325430$ & $6.126970300$
      \\ \hline
      $\gamma_e$ & $0.499$ & $0.5$ & $0.4$ \\
      $\gamma_\mu$ & $0.5$ & $-0.5$ & $-0.34$ \\
      $\gamma_\tau$ & $-0.5$ & $0.05$ & $0.49$ \\ \hline
      BR$\left( \mu^- \to e^- e^+ e^- \right)$ & $4.9 \times 10^{-13}$
      & $4.0 \times 10^{-13}$ & $3.0 \times 10^{-13}$ \\
      BR$\left( \tau^- \to e^- e^+ e^- \right)$ & $1.1 \times 10^{-9}$
      & $2.4 \times 10^{-8}$ & $2.5 \times 10^{-9}$ \\
      BR$\left( \tau^- \to e^- \mu^+ \mu^- \right)$ & $1.1 \times 10^{-9}$
      & $2.4 \times 10^{-8}$ & $1.9 \times 10^{-9}$ \\
      BR$\left( \tau^- \to \mu^- \mu^+ \mu^- \right)$ &
      $1.5 \times 10^{-8}$ & $1.0 \times 10^{-9}$ &
      $1.1 \times 10^{-9}$ \\ 
      BR$\left( \tau^- \to \mu^- e^+ e^- \right)$ &
      $1.5 \times 10^{-8}$ & $1.0 \times 10^{-9}$ &
      $1.5 \times 10^{-9}$ \\ \hline
      $X_{\mu e}^\mathrm{(loop)} / X_{\mu e}^\mathrm{(tree)}$ &
      $-2.69$ & $-1/505$ & $1/1020$ \\
      $X_{\tau e}^\mathrm{(loop)} / X_{\tau e}^\mathrm{(tree)}$ &
      $8.02$ & $20.3$ & $1/21.2$ \\
      $X_{\tau \mu}^\mathrm{(loop)} / X_{\tau \mu}^\mathrm{(tree)}$ &
      $7.93$ & $1.33$ & $1/1.39$ \\ \hline
    \end{tabular}
  \end{center}
  \caption{Three benchmark points. \label{bench}}
\end{table}
The first nine lines of that table contain the input to~\eqref{uit5},
\textit{viz.}\ the matrices $\Delta_1$ and $\Delta_2$,
the lightest-neutrino mass $m_1$,
and the new-scalar masses $M_3$ and $M_4$.
In the next six lines of table~\ref{bench} one sees the output of~\eqref{uit5},
\textit{viz.}\ the heavy-neutrino masses $\epsilon_{3+j}^2 m_{3+j}$
and the angles $\theta_{ij}^R$ that parameterize the matrix $U_R$.
In the next three lines of table~\ref{bench}
one finds the parameters $\gamma_\ell$
that we have fitted in order to obtain the desirable branching ratios
which are in the ensuing five lines of the table.
Finally,
in the last three lines of table~\ref{bench}
we compare the values of the quantities $X_{\ell_1 \ell_2}^\mathrm{(loop)}$
that were obtained from the solution of the one-loop equation~\eqref{uit5}
to the quantities $X_{\ell_1 \ell_2}^\mathrm{(tree)}$ that result
from the solution to the tree-level equation~\eqref{uit6}.

All the points in table~\ref{bench} have small $A_2$ asymmetries.
The asymmetry $A_1$ is also small for point~1,
but not for points~2 and~3;
the latter points rely on very small $X_{\mu e}^\mathrm{(loop)}$
to suppress BR$\left( \mu^- \to e^- e^+ e^- \right)$.

\section{Conclusions}
\label{conclusions}

The predictions for the lepton-flavour-violating charged-lepton decays
may be used to discriminate among theoretical models. 
For instance,
it has been found~\cite{novales} that,
in a model with a heavy charged gauge boson,
the present bounds on $\mu^- \to e^- \gamma$ and $\mu^- \to e^- e^+ e^-$
restrict the parameters of the model in such a way that the decays
$\tau^- \to \ell_2^- \ell_3^+ \ell_3^-$
($\ell_2, \ell_3 = e, \mu$)
will be invisible in the foreseeable future.
In this letter we have considered a model with radically different predictions.
In our model,
LFV decays like $\mu^- \to e^- \gamma$ and $Z \to e^+ \mu^-$ are invisible,
while $\mu^- \to e^- e^+ e^-$ and $\tau^- \to \ell_2^- \ell_3^+ \ell_3^-$
might be observed in future experiments.

Here we have investigated the one-loop radiative corrections
to the light-neutrino mass matrix
and their impact on the branching ratios 
$\mbox{BR} \left( \ell_1^- \to \ell_2^- \ell_3^+ \ell_3^- \right)$.
That impact occurs because the radiative corrections
strongly influence the evaluation of the heavy-neutrino masses $m_{4,5,6}$ 
and of the mixing matrix $U_R$ of the heavy neutrinos.
In our model $\mbox{BR} \left( \ell_1^- \to \ell_2^- \ell_3^+ \ell_3^- \right)$
is proportional to $\left| X_{\ell_1 \ell_2} \right|^2$,
where the quantities $X_{\ell_1 \ell_2}$ defined in~\eqref{X}
depend on $m_{4,5,6}$ and $U_R$.

We have shown that the one-loop radiative corrections
to the light-neutrino mass matrix may modify that matrix so much
that the model's predictions for $\mu^- \to e^- e^+ e^-$
and $\tau^- \to \ell_2^- \ell_3^+ \ell_3^-$ change drastically.
This is especially true for $\mbox{BR} \left( \mu^- \to e^- e^+ e^- \right)$,
which may shift by several orders of magnitude
when one (dis)considers the effect of the radiative corrections
on the determination of the heavy-neutrino masses and mixings.
This happens,
in particular,
because $X_{\mu e}$ may be zero for different values of the model's parameters
at the tree level and at the one-loop level.
The predictions for the four decays $\tau^- \to \ell_2^- \ell_3^+ \ell_3^-$
usually change by no more than two orders of magnitude
when one takes into account the radiative corrections,
but for values of the Yukawa couplings
%%%%% larger than the ones displayed in figure~\ref{fig4} the effects on
larger than the ones displayed in figure~\ref{fig4}
and given in~\eqref{vmdlfo0},
the effects on
$\mbox{BR} \left( \tau^- \to \ell_2^- \ell_3^+ \ell_3^- \right)$
may be dramatic too.

Our work highlights the necessity of taking into account
the one-loop radiative corrections to the light-neutrino mass matrix
when making any numerical assessment or prediction
of an effect that involves the masses $m_{4,5,6}$ and the mixing matrix $U_R$.
Usage of the standard seesaw formula~\eqref{vfpro}
is not adequate when one looks for detailed numerical predictions
because the `scotogenic-type' contributions to $\delta M_L$
in~\eqref{uit3} and~\eqref{uit4} may be non-negligible or even dominant.
This happens even when one takes into account
the restrictions posed by unitarity of the scalar potential
on the squared-mass differences among the neutral scalars;
though those differences are rather small,
the effects of the radiative corrections are nevertheless large in general.

\vspace*{5mm}

\paragraph{Acknowledgements:}
E.H.A.\ was supported partly by the FWF Austrian Science Fund
under the Doctoral Program W1252-N27 ``Particles and Interactions.''
P.M.F.\ is supported by
\textit{Funda\c c\~ao para a Ci\^encia e a Tecnologia} (FCT)
through
contracts
UIDB/00618/2020 and UIDP/00618/2020
and by HARMONIA project's contract UMO-2015/18/M/ST2/00518.
D.J.\ thanks the Lithuanian Academy of Sciences
for support through project DaFi2019.
Both P.M.F.\ and L.L.\ are supported by FCT project CERN/FIS-PAR/0004/2019.
L.L.\ has financial support of FCT through projects CERN/FIS-PAR/0008/2019,
PTDC/FIS-PAR/29436/2017,
UIDB/00777/2020,
and UIDP/00777/2020.

\newpage

\begin{appendix}

\setcounter{equation}{0}
\renewcommand{\theequation}{A\arabic{equation}}

\section{The maximum possible value of $\left| M_3^2 - M_4^2 \right|$}

\label{appendixA}

In this appendix we study in detail the scalar potential
of the 2HDM with alignment.
Our purpose is to demonstrate that
the difference between the squared masses
of the two new neutral scalars of that model may reach
$v^2 \left( 8 \pi / 3 \right) \approx 5.07 \times 10^5\,\mathrm{GeV}^2$.
We do not claim this to be an absolute upper bound;
simply,
we were able to demonstrate analytically that it may be reached.
On the other hand,
numerical scans that two of us have performed~\cite{nosso}
suggest that $8 \pi / 3$ is indeed the maximum possible value
of the parameter $\lambda_5$ of the scalar potential,
even in the general case without alignment.

\subsection{The scalar potential of the 2HDM}

Let the two doublets be
$\Phi_1 = \left( \begin{array}{cc}
  \varphi_1^+, & \varphi_1^0 \end{array} \right)^T$
and $\Phi_2 = \left( \begin{array}{cc}
  \varphi_2^+, & \varphi_2^0 \end{array} \right)^T$.
The scalar potential is~\cite{review}
\bs
\ba
V &=&
\mu_1\, \Phi_1^\dagger \Phi_1 + \mu_2\, \Phi_2^\dagger \Phi_2
+ \left( \mu_3\, \Phi_1^\dagger \Phi_2 + \mathrm{H.c.} \right)
\\ & &
+ \frac{\lambda_1}{2} \left( \Phi_1^\dagger \Phi_1 \right)^2
+ \frac{\lambda_2}{2} \left( \Phi_2^\dagger \Phi_2 \right)^2
+ \lambda_3\, \Phi_1^\dagger \Phi_1\, \Phi_2^\dagger \Phi_2
+ \lambda_4\, \Phi_1^\dagger \Phi_2\, \Phi_2^\dagger \Phi_1
\\ & &
+ \left[ \frac{\lambda_5}{2} \left( \Phi_1^\dagger \Phi_2 \right)^2
  + \left( \lambda_6\, \Phi_1^\dagger \Phi_1
  + \lambda_7\, \Phi_2^\dagger \Phi_2 \right) \Phi_1^\dagger \Phi_2
  + \mathrm{H.c.} \right],
\ea
\es
where $\mu_{1,2}$ and $\lambda_{1, \ldots, 4}$ are real
while $\mu_3$ and $\lambda_{5, \ldots, 7}$ are in general complex.
It is convenient to define
\be
\lambda_\pm := \frac{\lambda_1 \pm \lambda_2}{2}
\quad \mbox{and} \quad
\bar \lambda_\pm := \lambda_6 \pm \lambda_7.
\ee

The coefficients $\lambda_{1, \ldots, 7}$ are subject to two types of conditions:
the unitarity conditions and the boundedness-from-below (BFB) conditions.

\subsection{Unitarity conditions}

We consider three matrices:
\bs
\ba
\mathcal{M}_1 &=& \left( \begin{array}{cccc}
  \lambda_+ + \lambda_4
  &
  \mathrm{Re}\, \bar \lambda_+
  &
  - \mathrm{Im}\, \bar \lambda_+
  &
  \lambda_-
  \\
  \mathrm{Re}\, \bar \lambda_+
  &
  \lambda_3 + \mathrm{Re}\, \lambda_5
  &
  - \mathrm{Im}\, \lambda_5
  &
  \mathrm{Re}\, \bar \lambda_-
  \\
  - \mathrm{Im}\, \bar \lambda_+
  &
  - \mathrm{Im}\, \lambda_5
  &
  \lambda_3 - \mathrm{Re}\, \lambda_5
  &
  - \mathrm{Im}\, \bar \lambda_-
  \\
  \lambda_-
  &
  \mathrm{Re}\, \bar \lambda_-
  &
  - \mathrm{Im}\, \bar \lambda_-
  &
  \lambda_+ - \lambda_4
\end{array}
\right),
\\
\mathcal{M}_2 &=& \left( \begin{array}{cccc}
  3 \lambda_+ + 2 \lambda_3 + \lambda_4
  &
  3\, \mathrm{Re}\, \bar \lambda_+
  &
  - 3\, \mathrm{Im}\, \bar \lambda_+
  &
  3 \lambda_-
  \\
  3\, \mathrm{Re}\, \bar \lambda_+
  &
  \lambda_3 + 2 \lambda_4 + 3\, \mathrm{Re}\, \lambda_5
  &
  - 3\, \mathrm{Im}\, \lambda_5
  &
  3\, \mathrm{Re}\, \bar \lambda_-
  \\
  - 3\, \mathrm{Im}\, \bar \lambda_+
  &
  - 3\, \mathrm{Im}\, \lambda_5
  &
  \lambda_3 + 2 \lambda_4 - 3\, \mathrm{Re}\, \lambda_5
  &
  - 3\, \mathrm{Im}\, \bar \lambda_-
  \\
  3 \lambda_-
  &
  3\, \mathrm{Re}\, \bar \lambda_-
  &
  - 3\, \mathrm{Im}\, \bar \lambda_-
  &
  3 \lambda_+ - 2 \lambda_3 - \lambda_4
\end{array}
\right),
\no & &
\\
\mathcal{M}_3 &=& \left( \begin{array}{ccc}
  \lambda_1 & \lambda_5 & \sqrt{2}\, \lambda_6 \\
  \lambda_5^\ast & \lambda_2 & \sqrt{2}\, \lambda_7^\ast \\
  \sqrt{2}\, \lambda_6^\ast & \sqrt{2}\, \lambda_7 & \lambda_3 + \lambda_4
\end{array} \right).
\ea
\es
The unitarity conditions are
the following~\cite{Kanemura:1993hm,Akeroyd:2000wc,Ginzburg:2005dt}:
the moduli of the eigenvalues of $\mathcal{M}_1$,
$\mathcal{M}_2$,
and $\mathcal{M}_3$,
and also $\left| \lambda_3 - \lambda_4 \right|$,
must be smaller than $8 \pi$.

\subsection{BFB conditions}

We consider the matrix
\be
\Lambda_E = \left( \begin{array}{cccc}
  \lambda_+ + \lambda_3
  &
  \mathrm{Re}\, \bar \lambda_+
  &
  - \mathrm{Im}\, \bar \lambda_+
  &
  \lambda_-
  \\
  - \mathrm{Re}\, \bar \lambda_+
  &
  - \lambda_4 - \mathrm{Re}\, \lambda_5
  &
  \mathrm{Im}\, \lambda_5
  &
  - \mathrm{Re}\, \bar \lambda_-
  \\
  \mathrm{Im}\, \bar \lambda_+
  &
  \mathrm{Im}\, \lambda_5
  &
  - \lambda_4 + \mathrm{Re}\, \lambda_5
  &
  \mathrm{Im}\, \bar \lambda_-
  \\
  - \lambda_-
  &
  - \mathrm{Re}\, \bar \lambda_-
  &
  \mathrm{Im}\, \bar \lambda_-
  &
  - \lambda_+ + \lambda_3
\end{array}
\right).
\ee
Let $\Lambda_0$,
$\Lambda_1$,
$\Lambda_2$,
and $\Lambda_3$ be the eigenvalues of $\Lambda_E$.
The BFB conditions are the following~\cite{Deshpande:1977rw,Maniatis:2006fs,
  Ivanov:2006yq,Ivanov:2007de,Silva}:
\begin{enumerate}
\item $\Lambda_0$,
  $\Lambda_1$,
  $\Lambda_2$,
  and $\Lambda_3$ are real.
\item The largest eigenvalue,
  say $\Lambda_0$,
  is positive.
\item The $\left( 1,\, 1 \right)$ matrix element of the $4 \times 4$ matrix
  \be
  \left( \Lambda_E - \Lambda_1 \times \mathbbm{1}_{4 \times 4} \right)
  \times
  \left( \Lambda_E - \Lambda_2 \times \mathbbm{1}_{4 \times 4} \right)
  \times
  \left( \Lambda_E - \Lambda_3 \times \mathbbm{1}_{4 \times 4} \right)
  \ee
  is positive.
\end{enumerate}

\subsection{The Higgs basis and the alignment limit}

Let $v = 246$\,GeV be the vacuum expectation value (VEV).
We use the Higgs basis and write $\Phi_{1,2}$ as in~\eqref{doublets}.
In order that the doublet $\Phi_2$ has no VEV,
the parameter $\mu_3$ must be equal to
$- \lambda_6 v^2 / 2$~\cite{lavourasilva}.
Moreover,
$\mu_1 = - \lambda_1 v^2 / 2$
so that $v$ is the correct value of the VEV~\cite{lavourasilva}.

The mass terms of $H^+$,
$H$,
$S_3^0$,
and $S_4^0$ are given by
\be
V = \cdots + m_C^2\, H^+ H^-
+ \frac{1}{2} \left( \begin{array}{ccc} H, & S_3^0, & S_4^0 \end{array} \right)
M \left( \begin{array}{c} H \\ S_3^0 \\ S_4^0 \end{array} \right),
\ee
where $m_C^2 = \mu_2 + v^2 \lambda_3 / 2$ is the squared mass
of the physical charged scalar and~\cite{lavourasilva}
\be
\label{ccuvirt}
M = \left( \begin{array}{ccc}
  v^2 \lambda_1
  &
  v^2\, \mathrm{Re}\, \tilde \lambda_6
  &
  - v^2\, \mathrm{Im}\, \tilde \lambda_6
  \\
  v^2\, \mathrm{Re}\, \tilde \lambda_6
  &
  m_C^2
  + v^2 \left. \left( \lambda_4
  + \mathrm{Re}\, \tilde \lambda_5 \right) \right/ 2
  &
  - v^2 \left. \mathrm{Im}\, \tilde \lambda_5 \right/ 2
  \\
  - v^2\, \mathrm{Im}\, \tilde \lambda_6
  &
  - v^2 \left. \mathrm{Im}\, \tilde \lambda_5 \right/ 2
  &
  m_C^2
  + v^2 \left. \left( \lambda_4
  - \mathrm{Re}\, \tilde \lambda_5 \right) \right/ 2
\end{array} \right),
\ee
where $\tilde \lambda_5 := e^{- 2 i \alpha} \lambda_5$
and $\tilde \lambda_6 := e^{- i \alpha} \lambda_6$.

We now assume \emph{alignment},
which means that $H$ has mass $m_H = 125$\,GeV
and does not mix with $S_3^0$ and $S_4^0$.
Clearly,
from~\eqref{ccuvirt},
the absence of mixing means $\lambda_6 = 0$,
while $m_H^2 = v^2 \lambda_1$,
hence
\be
\label{l1}
\lambda_1 = \left( \frac{125}{246} \right)^2 \approx 0.258.
\ee
Alignment can be enforced through a $\mathbbm{Z}_2$ symmetry
$\Phi_2 \to - \Phi_2$ in the so-called inert 2HDM~\cite{Deshpande:1977rw,
  Barbieri:2006dq,Cao:2007rm,LopezHonorez:2006gr}.
However,
that possibility is not suitable for our purposes,
because we need all the Yukawa couplings in~\eqref{yukawas} to be nonzero.
Therefore,
in this letter alignment is just an \textit{ad hoc}\/ assumption.
We choose the phase $\alpha$ to offset $\arg{\lambda_5}$,
\textit{viz.}\ we choose
$e^{- 2 i \alpha} \lambda_5 = \pm \left| \lambda_5 \right|$.
Then,
from~\eqref{ccuvirt} the squared masses of $S_3^0$ and $S_4^0$ are
\be
M_3^2 = m_C^2 + v^2
\left. \left( \lambda_4 \pm \left| \lambda_5 \right| \right) \right/ 2
\quad \mbox{and} \quad
M_4^2 = m_C^2 + v^2
\left. \left( \lambda_4 \mp \left| \lambda_5 \right| \right) \right/ 2,
\ee
respectively.
Their difference is given by $\left| M_3^2 - M_4^2 \right|
= v^2 \left| \lambda_5 \right|$,
just as in the scotogenic model~\cite{ma}.
Thus,
finding the maximum possible value of $\left| M_3^2 - M_4^2 \right|$
is equivalent to finding the maximum possible value
of $\left| \lambda_5 \right|$,
which is determined by the unitarity and BFB conditions.

\subsection{Additional conditions}

One must guarantee that our assumed vacuum state
is indeed the state with the lowest value of $V$,
\textit{viz.}\ that we are not in the situation where
there are two local minima of the potential
and we are sitting on the local minimum
with the \emph{highest}\/ value of $V$ instead of being at the true vacuum;
this undesirable situation has been called `panic vacuum'.
This produces the following condition~\cite{Ivanov:2006yq,Ivanov:2007de,
  Barroso:2012mj,Silva}.
Let $\zeta \equiv 2 m_C^2 / v^2$ and let us order the eigenvalues of $\Lambda_E$
as $\Lambda_0 > \Lambda_1 > \Lambda_2 > \Lambda_3$.
Then,
either $\zeta > \Lambda_1$ or $\Lambda_2 > \zeta > \Lambda_3$.

There is also a phenomenological condition arising from the oblique parameter
$T$.
With alignment~\cite{osland},
\be
T = \frac{1}{16 \pi s_w^2 m_W^2} \left[
  f \left( m_C^2,\ M_3^2 \right)
  +
  f \left( m_C^2,\ M_4^2 \right)
  -
  f \left( M_3^2,\ M_4^2 \right)
  \right],
\ee
where $s_w^2 = 0.22$
is the squared sine of the weak mixing angle,
$m_W = 80.4$\,GeV is the mass of the $W^\pm$ gauge bosons,
and
\be
f \left( a,\ b \right) = \left\{ \begin{array}{lcl}
  {\displaystyle \frac{a + b}{2} - \frac{a b}{a - b}\, \ln{\frac{a}{b}}}
  & \Leftarrow & a \neq b,
  \\*[3mm]
  {\displaystyle 0} & \Leftarrow & a = b.
  \end{array} \right.
\ee
The phenomenological constraint is $T = 0.03 \pm 0.12$~\cite{rpp}.

\subsection{The special case $\lambda_1 = \lambda_2,\
  \lambda_6 = \lambda_7 = 0$}

When $\lambda_6 = \lambda_7 = 0$,
\textit{i.e.}\ $\bar \lambda_+ = \bar \lambda_- = 0$,
the matrices $\mathcal{M}_{1,2,3}$ and $\Lambda_E$
decompose as $2 \times 2$ matrices,
their eigenvalues are easy to compute,
and the unitarity and BFB conditions become much simpler~\cite{review}.
With the additional condition $\lambda_1 = \lambda_2$,
they are
\bs
\label{unit}
\ba
\left| \lambda_3 \pm \lambda_4 \right| &<& 8 \pi,
\\
\left| \lambda_3 \pm \left| \lambda_5 \right| \right| &<& 8 \pi,
\\
\left| \lambda_3 + 2 \lambda_4 \pm 3 \left| \lambda_5 \right| \right| &<& 8 \pi,
\\
\left| \lambda_1 \pm \left| \lambda_5 \right| \right| &<& 8 \pi,
\\
\left| \lambda_1 \pm \lambda_4 \right| &<& 8 \pi,
\\
\left| 2 \lambda_3 + \lambda_4 \pm 3 \lambda_1 \right| &<& 8 \pi,
\\
\lambda_1 &>& 0,
\\
\lambda_3 &>& - \lambda_1,
\\
\left| \lambda_5 \right| &<& \lambda_1 + \lambda_3 + \lambda_4.
\ea
\es
Notice that in this case
\be
\Lambda_0 = \lambda_3 + \lambda_1
\ee
while $\Lambda_1$,
$\Lambda_2$,
and $\Lambda_3$ are some permutation of
\be
\lambda_3 - \lambda_1,
\quad
- \lambda_4 + \left| \lambda_5 \right|,
\quad \mbox{and} \
- \lambda_4 - \left| \lambda_5 \right|.
\ee

\subsection{A solution}

With $\lambda_1$ given by~\eqref{l1},
there is a solution to~\eqref{unit}:
\be
\label{solution}
\lambda_3 = \frac{16 \pi}{3} - 2 \lambda_1 - \epsilon,
\quad \quad
\lambda_4 = - \frac{8 \pi}{3} + \lambda_1 + \epsilon,
\quad \quad
\left| \lambda_5 \right| = \frac{8 \pi}{3} - \epsilon,
\ee
where
\be
0 < \epsilon < \frac{8 \pi}{3}.
\ee
With this solution we learn that
$\left| \lambda_5 \right|$ may be as high as $8.377$,
and therefore
$\left| M_3^2 - M_4^2 \right| \lesssim 5.07 \times 10^5\,\mathrm{GeV}^2$.
For instance,
with $\left. \left( M_3 + M_4 \right) \right/ \! 2 = 1$\,TeV
one has $\left| M_3 - M_4 \right| \lesssim 253$\,GeV.

With~\eqref{solution},
\be
\label{fmddpp}
\left\{ M_3^2,\ M_4^2 \right\} = \left\{ m_C^2 + v^2\, \frac{\lambda_1}{2},\
m_C^2 + v^2
\left( - \frac{8\pi}{3} + \frac{\lambda_1}{2} + \epsilon \right) \right\}.
\ee
We are interested in the situation where $\epsilon$ is rather small,
so that $\left| \lambda_5 \right|$ is not very far from $8 \pi / 3$.
When $\epsilon$ is small,
$m_C^2$ lies in between the $M_3^2$ and $M_4^2$ given in~\eqref{fmddpp},
but it is very close to one of them because $\lambda_1$ is so small.
Then,
$T$ is negative but very small,
automatically satisfying the phenomenological constraint on that parameter.

With~\eqref{solution} one has
\be
\left\{ \Lambda_1,\ \Lambda_2 \right\} =
\left\{
\frac{16 \pi}{3} - 3 \lambda_1 - \epsilon,\
\frac{16 \pi}{3} - \lambda_1 - 2 \epsilon \right\}.
\ee
Therefore,
if we choose
\be
\label{limbound}
m_C^2 \ge v^2 \left( \frac{8 \pi}{3} - \frac{\lambda_1}{2} \right),
\ee
we avoid the undesirable situation of panic vacuum.
Thus,
we must have $m_C \gtrsim 707$\,GeV.
This lower bound on $m_C$ coincides with an analogous bound
obtained in a recent phenomenological analysis~\cite{analysis}.

Our solution~\eqref{solution} explicitly demonstrates that
$\left| M_3^2 - M_4^2 \right|$ may reach $v^2 \left( 8 \pi / 3 \right)$
without violating the unitarity and BFB conditions
and with a very small oblique parameter $T$.
Moreover,
the inequality~\eqref{limbound} provides a way
to choose the mass of the physical charged scalar
such as to evade panic vacuum.

\end{appendix}


\newpage


\begin{thebibliography}{99}

\bibitem{nobel1}
  Y.~Fukuda {\it et al.} [Super-Kamiokande Collaboration],
  \textit{Evidence for oscillation of atmospheric neutrinos},
  Phys.\ Rev.\ Lett.\ {\bf 81} (1998) 1562
  [{\tt hep-ex/9807003}].

\bibitem{nobel2}
  Q.~R.~Ahmad {\it et al.} [SNO Collaboration],
  \textit{Measurement of the rate of $\nu_e + d \to p + p + e^-$
  interactions produced by $^8\!B$ solar neutrinos
  at the Sudbury Neutrino Observatory},
  Phys.\ Rev.\ Lett.\ {\bf 87} (2001) 071301
  [{\tt nucl-ex/0106015}].

\bibitem{nobel3}
  B.~Aharmim {\it et al.} [SNO Collaboration],
  \textit{Combined analysis of all three phases of solar neutrino data
    from the Sudbury Neutrino Observatory},
  Phys.\ Rev.\ C {\bf 88} (2013) 025501
  [{\tt arXiv:1109.0763 [nucl-ex]}].

\bibitem{rpp}
  P.~A.~Zyla \textit{et al.}\ (Particle Data Group),
  \textit{The Review of Particle Physics},
  to be published in Prog.\ Theor.\ Exp.\ Phys.\ \textbf{2020} (2020) 083C01.

\bibitem{Mu3e}
  A.~Blondel \textit{et al.},
  \textit{Research proposal for an experiment
    to search for the decay $\mu \to eee$},
         {\tt arXiv:1301.6113 [physics.ins-det]}.

\bibitem{aushev}
  T.~Aushev \textit{et al.},
  \textit{Physics at Super $B$ Factory},
         {\tt arXiv:1002.5012 [hep-ex]}.

\bibitem{cerri}
  A.~Cerri \textit{et al.},
  \textit{Report from Working Group 4},
  CERN Yellow Rep.\ Monogr.\ \textbf{7} (2019) 867
  [{\tt arXiv:1812.07638 [hep-ph]}].

\bibitem{abdul}
  R.~Abdul~Khalek \textit{et al.},
  \textit{Standard Model Physics at the HL-LHC and HE-LHC},
  {\tt arXiv:1902.04070 [hep-ph]}.

\bibitem{Belle-II}
  W.~Altmannshofer \textit{et al.} [Belle-II Collaboration],
  \textit{The Belle II Physics Book},
  Prog.\ Theor.\ Exp.\ Phys.\ \textbf{2019} (2019) 123C01
  [{\tt arXiv:1808.10567 [hep-ex]}].

\bibitem{vicente}
  A.~Vicente,
  \textit{Higgs Lepton Flavor Violating Decays in Two Higgs Doublet Models},
  Front.\ Phys.\ \textbf{7} (2019) 174
  [{\tt arXiv:1908.07759 [hep-ph]}].

\bibitem{petcov}
  S.~T.~Petcov,
  \textit{The processes $\mu \to e \gamma$,
    $\mu \to e e e$,
    $\nu' \to \nu \gamma$ in the Weinberg--Salam Model with neutrino mixing},
  Sov.\ J.\ Nucl.\ Phys.\ \textbf{25} (1977) 340
  [erratum \textit{ibid.}\ \textbf{25} (1977) 698].

\bibitem{bilenky}
  S.~M.~Bilenky and S.~T.~Petcov,
  \textit{Massive neutrinos and neutrino oscillations},
  Rev.\ Mod.\ Phys.\ \textbf{59} (1987) 671
  [errata \textit{ibid.}\ \textbf{60} (1988) 575,
    \textbf{61} (1989) 169].

  %%%%% NEW REFERENCE
\bibitem{roig}
  G.~Hern\'andez-Tom\'e, G.~L\'opez Castro, and P.~Roig,
  \textit{Flavor violating leptonic decays of $\tau$ and $\mu$ leptons
    in the Standard Model with massive neutrinos},
  Eur.\ Phys.\ J.\ C \textbf{79} (2019) 84
  [erratum \textit{ibid.}\ \textbf{80} (2020) 438]
  [{\tt arXiv:1807.06050 [hep-ph]}].

\bibitem{blackstone}
  P.~Blackstone, M.~Fael, and E.~Passemar,
  \textit{$\tau \rightarrow \mu \mu \mu $
    at a rate of one out of $10^{14}$ tau decays?},
  Eur.\ Phys.\ J.\ C \textbf{80} (2020) 506
  [{\tt arXiv:1912.09862 [hep-ph]}].

\bibitem{GL1}
  W.~Grimus and L.~Lavoura,
  \textit{Soft lepton flavor violation in a multi-Higgs-doublet seesaw model},
  Phys.\ Rev.\ D \textbf{66} (2002) 014016
  [{\tt hep-ph/0204070}].

\bibitem{aeikens}
  E.~Aeikens, W.~Grimus, and L.~Lavoura,
  \textit{Charged-lepton decays from soft flavour violation},
  Phys.\ Lett.\ B \textbf{768} (2017) 365
  [{\tt arXiv:1612.00724 [hep-ph]}].

\bibitem{review}
  G.~C.~Branco, P.~M.~Ferreira, L.~Lavoura, M.~N.~Rebelo, M.~Sher,
  and J.~P.~Silva,
  \textit{Theory and phenomenology of two-Higgs-doublet models},
  Phys.\ Rept.\ {\bf 516} (2012) 1
  [{\tt arXiv:1106.0034 [hep-ph]}].

\bibitem{ivanov}
  I.~P.~Ivanov,
  \textit{Building and testing models with extended Higgs sectors},
  Prog.\ Part.\ Nucl.\ Phys.\ {\bf 95} (2017) 160
  [{\tt arXiv:1702.03776 [hep-ph]}].

\bibitem{seesaw1}
  P.~Minkowski,
  \textit{$\mu \to e \gamma$ at a rate of one out of $10^9$ muon decays?},
  Phys.\ Lett.\ \textbf{67B} (1977) 421.

\bibitem{seesaw2}
  T.~Yanagida,
  \textit{Horizontal gauge symmetry and masses of neutrinos},
  in \textit{Proceedings of the workshop on unified theory
    and baryon number in the universe (Tsukuba, Japan, 1979)},
  O.~Sawata and A.~Sugamoto eds.,
  KEK report \textbf{79-18},
  Tsukuba, 1979.

\bibitem{seesaw3}
  S.~L.~Glashow,
  \textit{The future of elementary particle physics},
  in \textit{Quarks and leptons,
    proceedings of the advanced study institute (Carg\`ese, Corsica, 1979)},
  M.~L\'evy \textit{et al.}~eds.,
  Plenum, New York, 1980.

\bibitem{seesaw4}
  M.~Gell-Mann, P.~Ramond, and R.~Slansky,
  \textit{Complex spinors and unified theories},
  in \textit{Supergravity},
  D.~Z.~Freedman and F.~van~Nieuwenhuizen eds.,
  North Holland, Amsterdam, 1979.

\bibitem{seesaw5}
  R.~N.~Mohapatra and G.~Senjanovi\'c,
  \textit{Neutrino mass and spontaneous parity violation},
  Phys.\ Rev.\ Lett.\ \textbf{44} (1980) 912.

\bibitem{soft}
  L.~Lavoura and W.~Grimus,
  \textit{Seesaw model with softly broken $L_e - L_\mu - L_\tau$},
  JHEP {\bf 0009} (2000) 007
  [{\tt hep-ph/0008020}].

\bibitem{chowdhury}
  T.~A.~Chowdhury and S.~Nasri,
  \textit{Charged lepton flavor violation
    in a class of radiative neutrino mass generation models},
  Phys.\ Rev.\ D \textbf{97} (2018) 075042
  [{\tt arXiv:1801.07199 [hep-ph]}].

\bibitem{GL2}
  W.~Grimus and L.~Lavoura,
  \textit{One-loop corrections to the seesaw mechanism
    in the multi-Higgs-doublet standard model},
  Phys.\ Lett.\ B \textbf{546} (2002) 86
  [{\tt hep-ph/0207229}].

\bibitem{neufeld}
  W.~Grimus and H.~Neufeld,
  \textit{Radiative neutrino masses in an $SU(2) \times U(1)$ model},
  Nucl.\ Phys.\ B \textbf{325} (1989) 18.

\bibitem{ibarra}
  A.~Ibarra and C.~Simonetto,
  \textit{Understanding neutrino properties
    from decoupling right-handed neutrinos and extra Higgs doublets},
  JHEP \textbf{1111} (2011) 022
  [{\tt arXiv:1107.2386 [hep-ph]}].

\bibitem{jurciukonis}
  D.~Jur\v{c}iukonis, T.~Gajdosik, and A.~Juodagalvis,
  \textit{Seesaw neutrinos
    with one right-handed singlet field and a second Higgs doublet},
  JHEP \textbf{1911} (2019) 146
  [{\tt arXiv:1909. 00752 [hep-ph]}].

\bibitem{aristizabal}
  D.~Aristizabal Sierra and C.~E.~Yaguna,
  \textit{On the importance of the 1-loop finite corrections
    to seesaw neutrino masses},
  JHEP \textbf{1108} (2011) 013
  [{\tt arXiv:1106.3587 [hep-ph]}].

\bibitem{osland}
  W.~Grimus, L.~Lavoura, O.~M.~Ogreid, and P.~Osland,
  \textit{A precision constraint on multi-Higgs-doublet models},
  J.\ Phys.\ G \textbf{35} (2008) 075001
  [{\tt arXiv:0711.4022 [hep-ph]}].

\bibitem{aad1}
  G.~Aad \textit{et al.} [ATLAS and CMS Collaborations],
  \textit{Combined Measurement of the Higgs Boson Mass
    in $pp$ Collisions at $\sqrt{s} = 7$ and 8\,TeV
    with the ATLAS and CMS Experiments},
  Phys.\ Rev.\ Lett.\ \textbf{114} (2015) 191803
  [{\tt arXiv:1503.07589 [hep-ex]}].

\bibitem{aad2}
  G.~Aad \textit{et al.} [ATLAS and CMS Collaborations],
  \textit{Measurements of the Higgs boson production and decay rates
    and constraints on its couplings from a combined ATLAS and CMS analysis
    of the LHC pp collision data at $\sqrt{s} = 7$ and 8\,TeV},
  JHEP \textbf{1608} (2016) 045
  [{\tt arXiv:1606.02266 [hep-ex]}].
  
\bibitem{ma}
 Ernest~Ma,
 \textit{Verifiable radiative seesaw mechanism of neutrino mass 
   and dark matter}, 
 Phys. Rev. D \textbf{73} (2006) 077301
 [{\tt hep-ph/0601225}].

\bibitem{esteban}
  I.~Esteban, M.~C.~Gonzalez-Garcia, A.~Hernandez-Cabezudo, M.~Maltoni,
  and T.\ Schwetz,
  \textit{Global analysis of three-flavour neutrino oscillations:
    synergies and tensions in the determination of $\theta_{23}$, $\delta_{CP}$,
    and the mass ordering},
  JHEP \textbf{1901} (2019) 106
  [{\tt arXiv:1811.05487 [hep-ph]}].

\bibitem{capozzi}
  F.~Capozzi, E.~Di Valentino, E.~Lisi, A.~Marrone, A.~Melchiorri,
  and A.~Palazzo,
  \textit{Global constraints on absolute neutrino masses and their ordering},
  Phys.\ Rev.\ D \textbf{95} (2017) 096014
  [erratum \textit{ibid.}\ \textbf{101} (2020) 116013]
  [{\tt arXiv:2003.08511 [hep-ph]}].

\bibitem{desalas}
  P.~F.~de Salas, D.~V.~Forero, S.~Gariazzo, P.~Mart\'\i nez-Mirav\'e,
  O.~Mena, C.~A.~Ternes, M.~T\'ortola, and J.~W.~F.~Valle,
  \textit{2020 Global reassessment of the neutrino oscillation picture},
         {\tt arXiv:2006.11237 [hep-ph]}.

\bibitem{esteban2020}
  I.~Esteban, M.~C.~Gonzalez-Garcia, M.~Maltoni, T.~Schwetz,
  and A.~Zhou,
  \textit{The fate of hints:
    updated global analysis of three-flavor neutrino oscillations},
         {\tt arXiv:2007.14792 [hep-ph]}.

\bibitem{Ade:2013zuv}
  Y.~Akrami {\it et al.} [Planck Collaboration],
  \textit{Planck 2018 results.\ I.\
  Overview and the cosmological legacy of Planck},
  {\tt arXiv:1807.06205 [astro-ph.CO]}.

\bibitem{analysis}
  D.~Chowdhury and O.~Eberhardt,
  \textit{Update of global Two-Higgs-Doublet model fits},
  JHEP {\bf 1805} (2018) 161
  [{\tt arXiv:1711.02095 [hep-ph]}].

\bibitem{nosso}
  D.~Jur\v{c}iukonis and L.~Lavoura,
  \textit{The three- and four-Higgs couplings
    in the general two-Higgs-doublet model},
  JHEP \textbf{1812} (2018) 004
  [{\tt arXiv:1807.04244 [hep-ph]}].

\bibitem{novales}
  H.~Novales-S\'anchez, M.~Salinas, and J.~J.~Toscano,
  \textit{About heavy neutrinos:
    Lepton-flavor violation in decays of charged leptons},
  J.\ Phys.\ G \textbf{45} (2018) 095004
  [{\tt arXiv:1710.08474 [hep-ph]}].

\bibitem{Kanemura:1993hm}
  S.~Kanemura, T.~Kubota, and E.~Takasugi,
  \textit{Lee--Quigg--Thacker bounds for Higgs boson masses
    in a two doublet model},
  Phys.\ Lett.\ B \textbf{313} (1993) 155
  [{\tt hep-ph/9303263}].

\bibitem{Akeroyd:2000wc}
  A.~G.~Akeroyd, A.~Arhrib, and E.~M.~Naimi,
  \textit{Note on tree level unitarity in the general two Higgs doublet model},
  Phys.\ Lett.\ B \textbf{490} (2000) 119
  [{\tt hep-ph/0006035}].

\bibitem{Ginzburg:2005dt}
  I.~F.~Ginzburg and I.~P.~Ivanov,
  \textit{Tree-level unitarity constraints in the most general 2HDM},
  Phys.\ Rev.\ D \textbf{72} (2005) 115010
  [{\tt hep-ph/0508020}].

\bibitem{Deshpande:1977rw}
  N.~G.~Deshpande and E.~Ma,
  \textit{Pattern of Symmetry Breaking with Two Higgs Doublets},
  Phys.\ Rev.\ D \textbf{18} (1978) 2574.

\bibitem{Maniatis:2006fs}
  M.~Maniatis, A.~von~Manteuffel, O.~Nachtmann, and F.~Nagel,
  \textit{Stability and symmetry breaking in the
    general two-Higgs-doublet model},
  Eur.\ Phys.\ J.\ C \textbf{48} (2006) 805
  [{\tt hep-ph/0605184}].

\bibitem{Ivanov:2006yq}
  I.~P.~Ivanov,
  \textit{Minkowski space structure of the Higgs potential in 2HDM},
  Phys.\ Rev.\ D \textbf{75} (2007) 035001
  [erratum \textit{ibid.}\ \textbf{76} (2007) 039902]
  [{\tt hep-ph/0609018}].

\bibitem{Ivanov:2007de}
  I.~P.~Ivanov,
  \textit{Minkowski space structure of the Higgs potential in 2HDM.
    II. Minima, symmetries, and topology},
  Phys.\ Rev.\ D \textbf{77} (2008) 015017
  [{\tt arXiv:0710.3490 [hep-ph]}].

\bibitem{Silva}
  I.~P.~Ivanov and J.~P.~Silva,
  \textit{Tree-level metastability bounds
    for the most general two Higgs doublet model},
  Phys.\ Rev.\ D {\bf 92} (2015) 055017
  [{\tt arXiv:1507.05100 [hep-ph]}].

\bibitem{lavourasilva}
  L.~Lavoura and J.~P.~Silva,
  \textit{Fundamental $CP$-violating quantities in an
    $SU(2) \times U(1)$ model with many Higgs doublets},
  Phys.\ Rev.\ D \textbf{50} (1994) 4619
  [{\tt hep-ph/9404276}].

\bibitem{Barbieri:2006dq}
  R.~Barbieri, L.~J.~Hall, and V.~S.~Rychkov,
  \textit{Improved naturalness with a heavy Higgs:
    An Alternative road to LHC physics},
  Phys.\ Rev.\ D \textbf{74} (2006) 015007
  [{\tt hep-ph/0603188}].

\bibitem{Cao:2007rm}
  Q.~H.~Cao, E.~Ma, and G.~Rajasekaran,
  \textit{Observing the Dark Scalar Doublet
    and its Impact on the Standard-Model Higgs Boson at Colliders},
  Phys.\ Rev.\ D \textbf{76} (2007) 095011
  [{\tt arXiv:0708.2939 [hep-ph]}].

\bibitem{LopezHonorez:2006gr}
  L.~Lopez Honorez, E.~Nezri, J.~F.~Oliver, and M.~H.~G.~Tytgat,
  \textit{The Inert Doublet Model: An Archetype for Dark Matter},
  JCAP \textbf{02} (2007) 028
  [{\tt arXiv:hep-ph/0612275 [hep-ph]}].

\bibitem{Barroso:2012mj}
  A.~Barroso, P.~M.~Ferreira, I.~P.~Ivanov, R.~Santos, and J.~P.~Silva,
  \textit{Evading death by vacuum},
  Eur.\ Phys.\ J.\ C \textbf{73} (2013) 2537
  [{\tt arXiv:1211.6119 [hep-ph]}].
  
\end{thebibliography}
\end{document}